# Synthesis of technetium hydride $TcH_{1.3}$ at 27 GPa

D. Zhou,[1,*] D. V. Semenok,[1,*] M. A. Volkov,[2] I. A. Troyan,[3] A. Yu. Seregin,[3,4] I. V. Chepkasov,[1] D. A. Sannikov,[5] P. G. Lagoudakis,[5] A. R. Oganov,[1] and K. E. German[2,*]

[1] Skolkovo Institute of Science and Technology, Bolshoy Boulevard 30, bld. 1, Moscow 121205, Russia
[2] A. N. Frumkin Institute of Physical Chemistry and Electrochemistry of the Russian Academy of Sciences, Department of Radiochemistry, 31 Leninsky Prospekt, Moscow 119991, Russia
[3] Shubnikov Institute of Crystallography, Federal Scientific Research Center Crystallography and Photonics, Russian Academy of Sciences, 59 Leninsky Prospekt, Moscow 119333, Russia
[4] National Research Center "Kurchatov Institute", Ploshchad' Akademika Kurchatova, 1, Moscow 123182, Russia
[5] Center for Photonics and Quantum Materials, Skolkovo Institute of Science and Technology, Bolshoy Boulevard 30, bld. 1, Moscow 121205, Russia

*Corresponding authors: d.zhou@skoltech.ru, dmitrii.semenok@skoltech.ru, guerman_k@mail.ru

## Abstract

In this work, we synthesize and investigate lower technetium hydrides at pressures up to 45 GPa using the synchrotron X-ray diffraction, reflectance spectroscopy, and ab initio calculations. In the Tc–H system, the hydrogen content in $TcH_x$ phases increases when the pressure rises, and at 27 GPa we found a new hexagonal (*hcp*) nonstoichiometric hydride $TcH_{1.3}$. The formation of technetium hydrides is also confirmed by the emergence of a new reflective band at 450–600 nm in the reflectance spectra of $TcH_x$ samples synthesized at 45 GPa. On the basis of the theoretical analysis, we proposed crystal structures for the phases $TcH_{0.45\pm0.05}$ ($TcH_{16}H_7$) and $TcH_{0.75\pm0.05}$ ($Tc_4H_3$) previously obtained at 1–2 GPa. The calculations of the electron–phonon interaction show that technetium hydrides $TcH_{1+x}$ do not possess superconducting properties due to the low electron–phonon interaction parameter ($\lambda \sim 0.23$).

## Introduction

The science and technology of polyhydride materials currently have two main vectors of development: materials for hydrogen storage and high-temperature superconductors. The first direction[1] recently received a significant boost with the discovery of pressure-stabilized polyhydrides with ultrahigh hydrogen content (up to 63 wt %) such as the molecular complexes of methane $(CH_4)_3(H_2)_{25}$,[2] hydrogen iodide $(HI)(H_2)_{13}$,[3] strontium and barium polyhydrides $SrH_{22}$[4] and $BaH_{12}$,[5] and various rubidium and cesium polyhydrides.[6] Although the stabilization of most polyhydrides currently requires considerable pressures of at least several gigapascals (Table 1), the conditions for their stabilization are continuously improving, which gives hope for a discovery of new stable hydrogen storage compounds with record capacity in the future.

**Table 1.** Hydrogen capacity of polyhydrides.

| Compound | Stabilization pressure, GPa | Hydrogen content, wt % |
|---|---|---|
| $CH_4$ | 0 | 25 |
| $NH_3BH_3$ | 0 | 19 |
| $RbH_{\sim9}$ | 8 | 9.5 |
| $CsH_{\sim17}$ | 10 | 11 |
| $BaH_{12}$ | 75 | 8 |
| $SrH_{22}$ | 80 | 20 |
| $Xe(H_2)_8$ | 5–8 | 11 |
| $(CH_4)_3(H_2)_{25} = CH_{20.7}$ | 10 | 63 |
| $(HI)(H_2)_{13}$ | 9 | 17 |

The second promising area of research in hydride chemistry is the designing of high-temperature superconductors with a near-room temperature of superconductivity.[7] At the moment, a well reproducible record superconductor is *fcc*-$LaH_{10}$ with a critical superconductivity temperature $T_C$ of $250 \pm 5$ K.[8–9] Superconductivity is distributed unevenly among polyhydrides, being a distinctive feature of mostly cubic and hexagonal hydrides such as $XH_6$, $XH_9$, $XH_{10}$, and tetragonal $XH_4$, with the atomic hydrogen sublattices stabilized at 100–200 GPa.[10–11] In this pressure range, the critical temperatures of the best superhydrides are in the range $T_C$ [K] $\subset (P, 2P)$ [GPa]. The hydride-forming element can be sulfur, an alkaline earth metal (Mg, Ca, Sr, Ba), or have only 1–2 d or f electrons (La, Y, Zr, Th, Ce, Lu, etc.). For all other elements of the periodic table (the "empty" zone), polyhydrides are either not formed or do not exhibit superconducting properties.

In this research, we investigate the Tc–H system, belonging to this "empty" zone, at pressures up to 45 GPa to verify theoretical predictions on the distribution of superconducting properties in polyhydrides.[12] Despite the obvious interest in hydrogen-saturated compounds like $K_2TcH_9$,[12] only a few papers on thermodynamic and quantum mechanical calculations of the Tc–H system for gaseous states of hydrides are found in the literature before 2021.[13–16] For example, the calculations of the enthalpy of formation of technetium monohydride[17] showed that $\Delta_f H°_m(½TcH) = +9$ kJ/mole, that is, TcH must decompose spontaneously at ambient pressure and can be stabilized only when the pressure is increased.

Experimental works in the synthesis of technetium hydrides at elevated pressures have shown that metallic technetium has a little tendency to react with $H_2$ under the ambient temperature and pressure conditions.[18] Spitsyn et al.[18] reported the synthesis of $TcH_{0.73\pm0.05}$ via a direct reaction of Tc with hydrogen gas at pressures up to 1.9 GPa and a temperature of 573 K. The resulting hydride formed a single phase with a hexagonal lattice: $a = 2.805 \pm 0.02$ Å, $c = 4.455 \pm 0.02$ Å. At an elevated pressure (2.2 GPa), the same team obtained two hexagonal phases with the compositions $TcH_{0.5}$ ($\varepsilon_1$) and $TcH_{0.78}$ ($\varepsilon_2$).[19] The $\varepsilon_1 \rightarrow \varepsilon_2$ phase transition, observed at about 1 GPa, was confirmed by resistivity measurements.

The formation of nonstoichiometric hydrides $TcH_{0.45}$, $TcH_{0.69}$,[20–21] and $TcH_m$, where $m = 0.26$,[22] 0.385, 0.485, and 0.765,[23] has been proved using the neutron powder diffraction. The hydrides with $m = 0.45$ and 0.69 have been indexed in the hexagonal space group *hcp*: $a = 2.801 \pm 0.004$ Å, $c = 4.454 \pm 0.01$ Å for $TcH_{0.45}$; $a = 2.838 \pm 0.004$ Å, $c = 4.465 \pm 0.01$ Å for $TcH_{0.69}$. Three unexpected peaks in the neutron diffraction patterns have been observed for $TcH_{0.45}$ and interpreted as the evidence of a superstructure. The authors concluded that hydrogen atoms occupy octahedral spaces in the technetium lattice. [20–23]

The studies of the transport properties and superconductivity of the obtained lower hydrides showed that hydrogen is highly negative to the electron–phonon interaction in technetium. The critical temperature $T_C$ of the pure metal is about 7.73 K at normal pressure, which is one of the highest values among the pure elements.[24] However, $TcH_{0.73\pm0.05}$ ($\varepsilon_2$) does not have a transition to the superconducting state above 2 K, probably because of the increasing Tc–Tc distance.[18,20,25–26]

Theoretical studies of stable compounds in the Tc–H system have been previously conducted using the density functional theory (DFT) methods at high and ultrahigh pressures up to 300 GPa.[27] It has been found that increasing the pressure leads to the formation of new polyhydrides: $P6_3/mmc$-TcH (stable in the range of 0–200 GPa), $I4/mmm$-$TcH_2$ (stable above 64 GPa), $Pnma$-$TcH_3$ (stable above 79 GPa), and $P4_2/mmc$-$TcH_3$ (forms at about 300 GPa), in which the charge is partially transferred from the technetium atoms to hydrogen.

The theoretically calculated superconducting properties of technetium hydrides are weakly expressed, with $T_C$ not exceeding 11 K.[27]

In this work, we synthesized and characterized technetium hydrides $TcH_{1+x}$ at pressures up to 27 GPa using the X-ray powder diffraction and at pressures up to 45 GPa using the reflection spectroscopy in diamond anvil cells (DACs).

## Results and Discussion

### *Structure search*

At present, theoretical analysis and evolutionary structural search are among the cornerstones in hydride chemistry[28–32] because the exact positions of hydrogen atoms in polyhydrides at high pressures cannot be established using experimental methods. We performed an evolutionary crystal structure search for thermodynamically stable phases at 5, 25, and 50 GPa in the Tc–H system using the USPEX code.[33–35] In terms of thermodynamics, the Tc–H system belongs to those where metal-rich phases dominate hydrogen-rich phases (e.g., Ba–H,[5] Sr–H,[4] etc.).

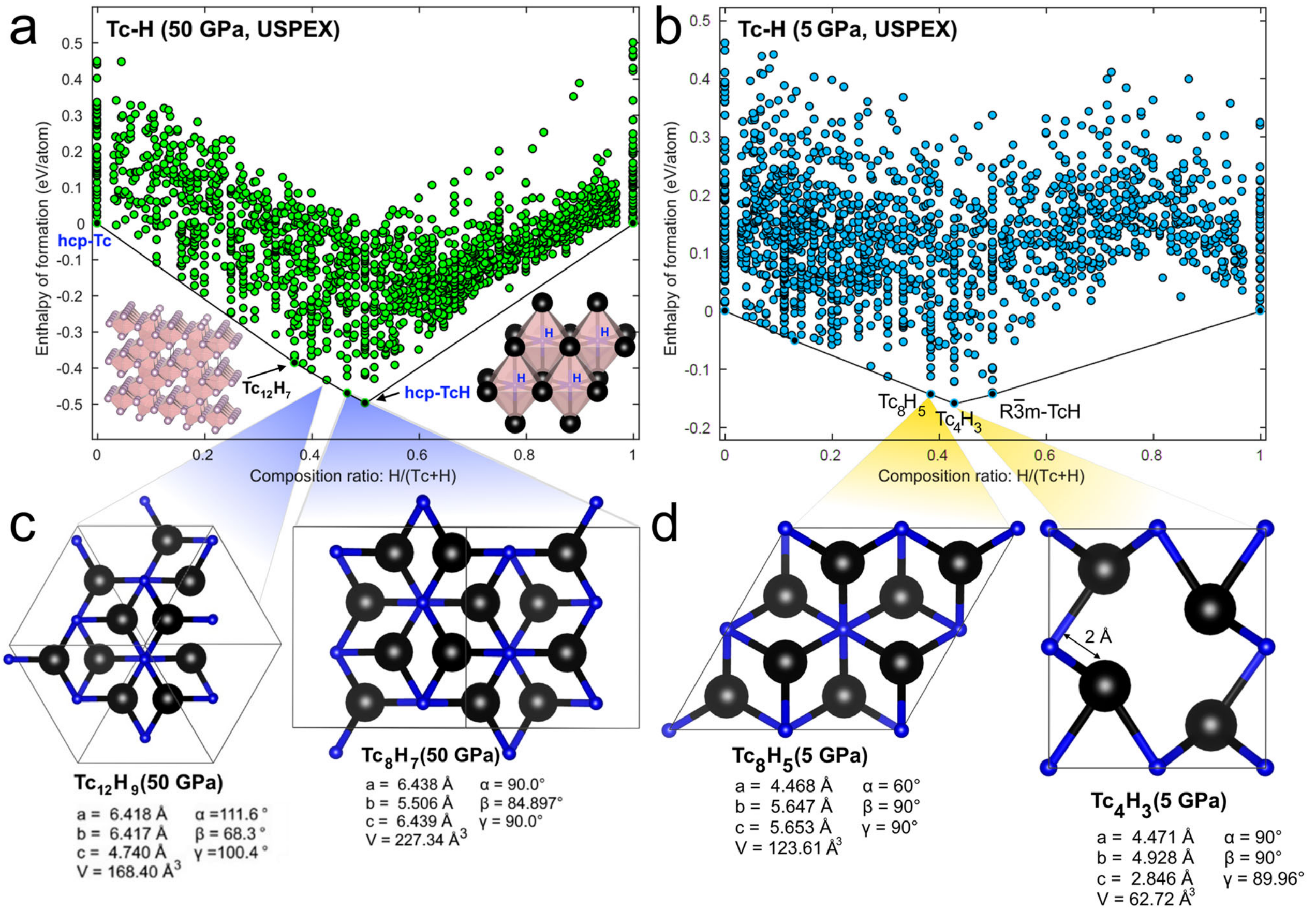

**Figure 1.** Calculated convex hulls for the Tc–H system at (a) 50 GPa, (b) 5 GPa, and (c, d) the most stable crystal structures of pseudohexagonal $TcH_x$ ($x = 0.75 \pm 0.12$). Inset: (pseudo)hexagonal structures of TcH and $Tc_{12}H_7$, where hydrogen (H) occupies octahedral voids. The predicted XRD patterns of these structures and their CIF files are given in the Supporting Information.

The calculations resulted in a phase diagram (convex hull, Figure 1) of technetium hydrides which shows that at 25 (see the Supporting Information, Figure S1) and 50 GPa the highest hydrogen content is achieved in $P6_3/mmc$-TcH, whereas all higher hydrides are thermodynamically unstable. Many nonstoichiometric $TcH_x$ hydrides ($x < 1$) lie close to the convex hull between Tc and TcH, which speaks in favor of the possible formation of hydrides with intermediate compositions ($0 < x < 1$) at low pressures or hydrogen deficiency. The most stable phases at 50 GPa are pseudohexagonal $P1$-$Tc_{12}H_7$, $P1$-$Tc_{12}H_9$, and $P1$-$Tc_8H_7$ (Figure 1). At 5 GPa, the thermodynamically stable phases are $Tc_4H_3$ and $Tc_{16}H_7$, which are candidates for previously experimentally found compounds $TcH_{0.75\pm0.05}$ ($\varepsilon_2$) and $TcH_{0.5\pm0.05}$ ($\varepsilon_1$), respectively.

## *Experimental synthesis*

Technetium samples mixed with ammonia borane $NH_3BH_3$ were heated in DACs T1 and T2 by a pulsed laser for several hundred microseconds to 1500 K at pressures of 34 GPa (DAC T1, 300 μm culet) and 45 GPa (DAC T2, determined via the Raman signal of diamond[36]). After the laser heating, the pressure in DAC T1 dropped to 27 GPa. Low-pressure X-ray diffraction studies were carried out on a synchrotron source of the Kurchatov Institute (KISI-Kurchatov), station RKFM (λ = 0.62 Å, 20 keV, beam width was about 50 μm).

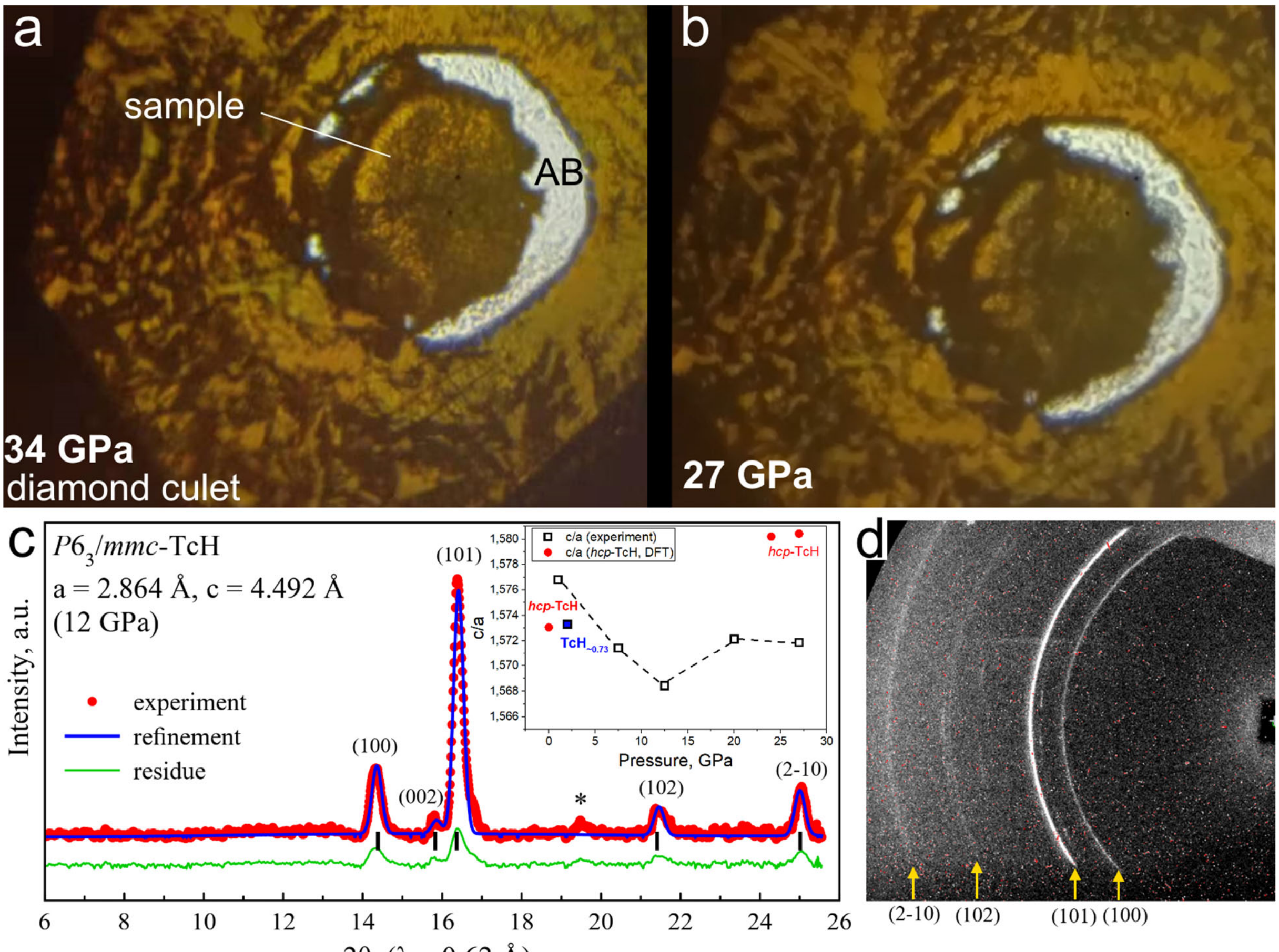

**Figure 2.** X-ray powder analysis of technetium hydrides in DAC T1. Microphotographs of the loaded Tc sample (a) before and (b) after the laser heating, where the pressure significantly decreased. (c) Experimental XRD pattern measured at 300 K and the Le Bail refinement of the unit cell parameters of $P6_3/mmc$-TcH at 12 GPa. The experimental data, fit, and residue are shown in red, blue, and green, respectively. The unidentified reflections are indicated by asterisks. Inset: pressure dependence of the $c/a$ ratio. (d) Experimental diffraction pattern of TcH.

Figure 2a,b shows a substantial change in the character of the technetium surface after the laser heating. As we will see below, the spectral reflectivity of the sample also changes significantly. The XRD analysis shows the presence of one hexagonal phase (Figure 2c), whose cell volume at pressures above 10 GPa slightly exceeds the values theoretically calculated for $P6_3/mmc$-TcH (Table 2), corresponding to compositions $TcH_{1.1\text{–}1.3}$. Some XRD patterns also show an *hcp* admixture of Re (or Tc), which is due to the large width of the X-ray beam (~50 μm). As the pressure in DAC T1 decreases, so does the hydrogen content in the compound, and at 1–2 GPa the cell parameters of $TcH_x$ approach the literature data for the composition $TcH_{0.73}$ (Figure 3).[18–19,23] This allows us to conclude that stoichiometric TcH is thermodynamically unstable below 10 GPa and loses hydrogen already at 300 K.

**Table 2.** Experimental unit cell parameters of synthesized technetium hydrides $TcH_{1+x}$ and metallic technetium ($V_{Tc}$).[37]

| Pressure, GPa | $a$, Å | $c$, Å | $c/a$ | $V$, Å$^3$ ($Z = 2$) | $V$, Å$^3$/Tc | $V_{Tc}$, Å$^3$/Tc |
|---|---|---|---|---|---|---|
| 27 | 2.845 | 4.472 | 1.572 | 31.35 | 15.67 | 13.28 |
| 20 | 2.849 | 4.479 | 1.572 | 31.49 | 15.74 | 13.51 |
| 12 | 2.864 | 4.492 | 1.568 | 31.91 | 15.95 | 13.79 |
| 7 | 2.856 | 4.488 | 1.571 | 31.70 | 15.85 | 14.00 |
| ~1 | 2.838 | 4.475 | 1.572 | 31.22 | 15.61 | 14.30 |
| ~2, ($TcH_{0.73}$) | 2.838 | 4.465 | 1.573 | 31.14 | 15.57 | 14.21 |
| ~2, ($TcH_{0.45}$) | 2.801 | 4.454 | 1.590 | 30.62 | 15.31 | |

The reaction of technetium with hydrogen at high pressures is very similar to the behavior of the Mo–H system, where, also at a pressure of about 15–20 GPa, a nonstoichiometric phase $MoH_{1.35}$ forms.[38] However, for example, the behavior of the Re–H system is different: at 20 GPa, only $ReH_{0.38}$, a hydride with substantially lower hydrogen content, is formed.[39] An estimation of the saturation composition of technetium hydride $TcH_{1+x}$ at 27 GPa is possible from volumetric considerations. Because the volume expansion per Tc atom is measured using X-ray diffraction, the composition H/Tc of the saturated Tc hydride can be estimated by comparing this volume expansion with the expected volume expansion per H atom. In the *hcp* structure, the occupation of all octahedral voids by hydrogen leads to TcH stoichiometry. When more hydrogen is used, it is only possible to place all extra hydrogens in the tetrahedral voids. Each hydrogen atom placed in an octahedral interstitial site in a close-packed lattice of technetium expands the lattice by about 1.86 Å$^3$ per H atom, whereas in a tetrahedral site the volume expansion is 2.2–3.2 Å$^3$ per H atom.[40] Considering that $\Delta V = V(TcH_{1+x}) - V(Tc) = 2.39$ (27 GPa), 2.23 (20 GPa), 2.16 (12 GPa), and 1.85 Å$^3$/Tc (7 GPa), we can estimate the maximum hydrogen content of the obtained hydride as $TcH_{1.3}$ (Figure 3a), which is quite close to the results of the first-principles calculations.

Hydrogen has a negative effect on the superconducting properties of technetium despite the fact that it only slightly directly affects the electronic structure of Tc hydrides, contributing almost nothing to the electron density of states at the Fermi level (Figure 4d, Supporting Information, Figures S6-7). This is due to the fact that hydrogen occupies the octahedral voids in the hexagonal lattice of technetium, increasing the parameters of the unit cell, which is equivalent to introducing a formally negative pressure to *hcp*-Tc. This concept is schematically shown in Figure 3d. For instance, an extrapolation of the equation of state of Tc (Figure 3a, Supporting Information Figure S3) to the negative pressure region shows that the cell volume of the synthesized hydride $TcH_{1+x}$ corresponds to a pressure of about –30 GPa. At the same time, the region of negative pressures from 0 to –30 GPa, where the unit cell of Tc is only slightly expanded, is unexplored, and a local increase in the critical temperature $T_C$ expected there requires either a compression of TcH above 50 GPa, or controlled hydrogenation with a very small amount of $H_2$. From a practical point of view, small levels of hydrogenation with a controlled expansion of the unit cell are achievable using the electrochemical approach.[41] At the same time, as the pressure increases from 0 to 1.5 GPa, the critical temperature of technetium decreases with a slope of $dT_C/dP = -0.125$ K/GPa.[42]

The Crystal Orbital Hamilton Population (COHP)[43] analysis shows (Supporting Information, Figure S10) that the technetium atoms interact quite strongly with each other (Tc-Tc) and with hydrogen (Tc-H, Figures S10c-f). However, in the vicinity of the Fermi level, the Tc-H interaction is virtually zero, which corresponds to the negligible contribution of the hydrogen sublattice to the conductivity and superconductivity of technetium hydrides. In the *hcp*-TcH there is a little electronic instability of the Tc-Tc bonds in the vicinity of the Fermi level (Figure S10a), as well as a pronounced contribution of the anti-bonding orbitals to the H-H interaction (Figure S10b). This may indicate that TcH will not be stable under decreasing pressure and will lose hydrogen.

According to the ab initio calculations (Figure 3c), $P6_3/mmc$-TcH does not have the superconducting properties because of low electron–phonon interaction ($\lambda_{SC} \approx 0.23$ at 30 GPa). In contrast, pure technetium at normal pressure has high $\lambda_{SC} \sim 0.8$ and $T_C = 7.7$ K,[44] possibly due to the proximity of the phase transition which is formally located in the "negative" pressure region. In the region of positive pressures, the hexagonal phase of technetium is stable at least up to 67 GPa.[37] These results are consistent with measurements done in the 1970s–80s, when it was shown that superconductivity in technetium is suppressed by both hydrogen and pressure.[18,20,25–26]

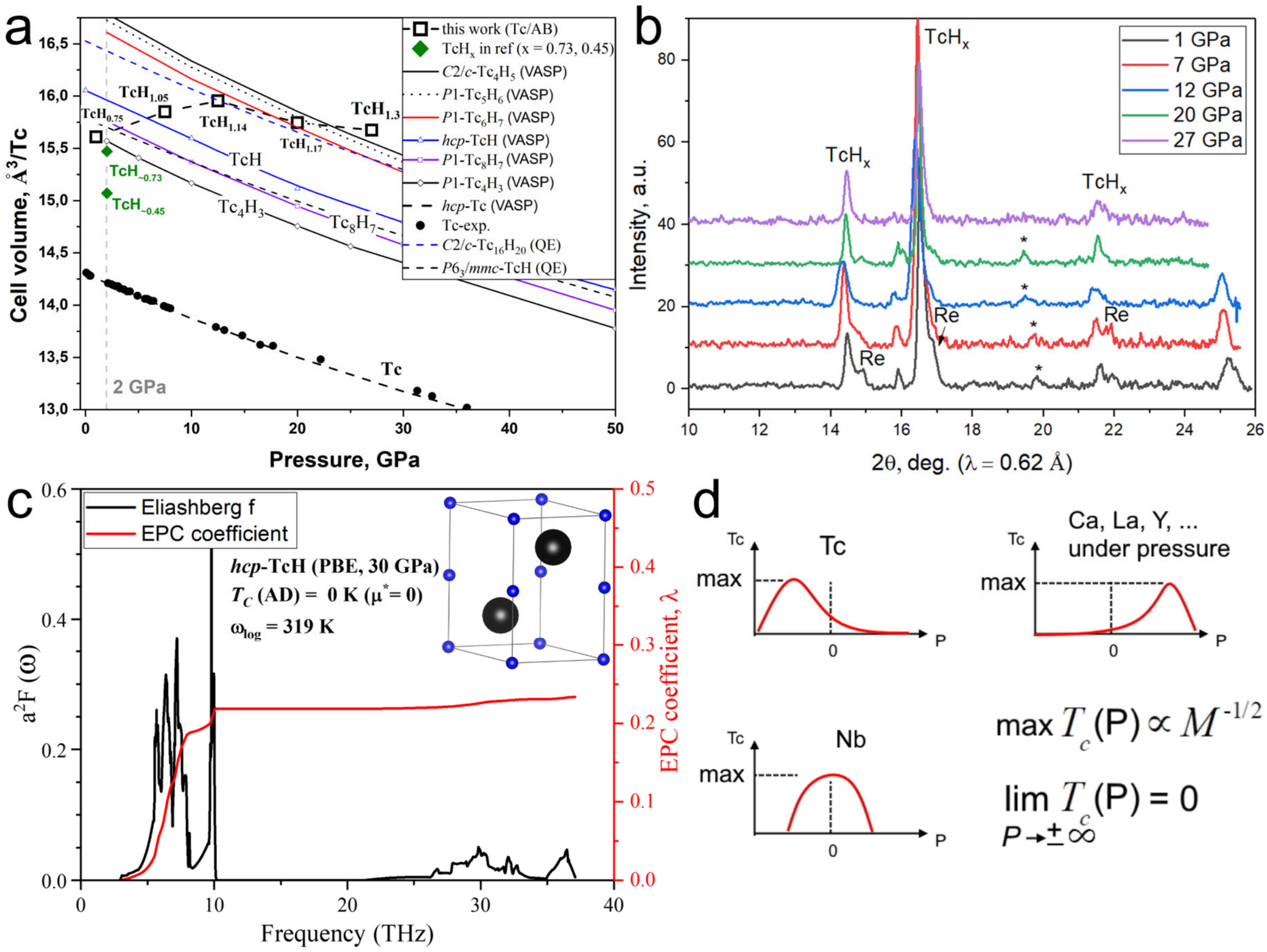

**Figure 3.** Physical properties of technetium hydrides. (a) Experimental and theoretical (VASP, QE) equations of state of various technetium hydrides and metallic technetium (see also Supporting Information). (b) X-ray diffraction patterns of the synthesized technetium hydrides during decompression from 27 to ~1 GPa. (c) Eliashberg function and the electron–phonon interaction parameter of technetium monohydride TcH at 30 GPa. The calculations were performed using the PBE–SP–HGH pseudopotentials in the Quantum Espresso code.[33] Inset: crystal structure of *hcp*-TcH. (d) Schematic representation of different variants of the pressure dependence of the critical temperature of conventional Bardeen–Cooper–Schrieffer superconductivity [45] for metals, where $M$ is the mass of an atom, $P$ is the pressure.

## *Reflectance spectroscopy*

We investigated the relative reflectance $R(\lambda)$ of metallic technetium and the synthesized $TcH_{1+x}$ hydrides in the high-pressure DAC T2 in the visible spectral region (400–900 nm, 1.4–3.1 eV, Figure 4). Because the geometry of the diamond anvil, sample, and $NH_3BH_3$ layer is unknown, it is only possible to determine the relative reflectivity of the sample in comparison with the reflection from either the Re gasket or from the empty place such as the diamond/$NH_3BH_3$ boundary (reference) of a diamond anvil (Figure 4c). Before starting the experiment, a series of calibrations were performed to determine the apparent spectral function of the incandescent lamp after passing the light through the optical system (Supporting Information Figures S16–S18). Then, the relative reflectance of the empty diamond cell, rhenium gasket, and copper and tungsten particles in ammonia borane media was measured at ambient pressure (Supporting Information Figures S19–S24). The complex interference pattern was smoothed (using the Fourier filter from OriginLab[46]) and multiplied by an arbitrary normalization constant. The relative reflectance values were calculated as $R(\lambda) = \text{const} \times I_{\text{sample}}(\lambda)/I_{\text{ref}}(\lambda)$, where $I_{\text{sample}}(\lambda)$ and $I_{\text{ref}}(\lambda)$ are intensities of light reflected from the sample and the reference, respectively. The obtained values agree with the known experimental data for the studied metals (Supporting Information Figures S25–S27). Reflectance spectra of the compressed sulfur hydride $H_3S$ were recently measured and processed in a similar manner [47].

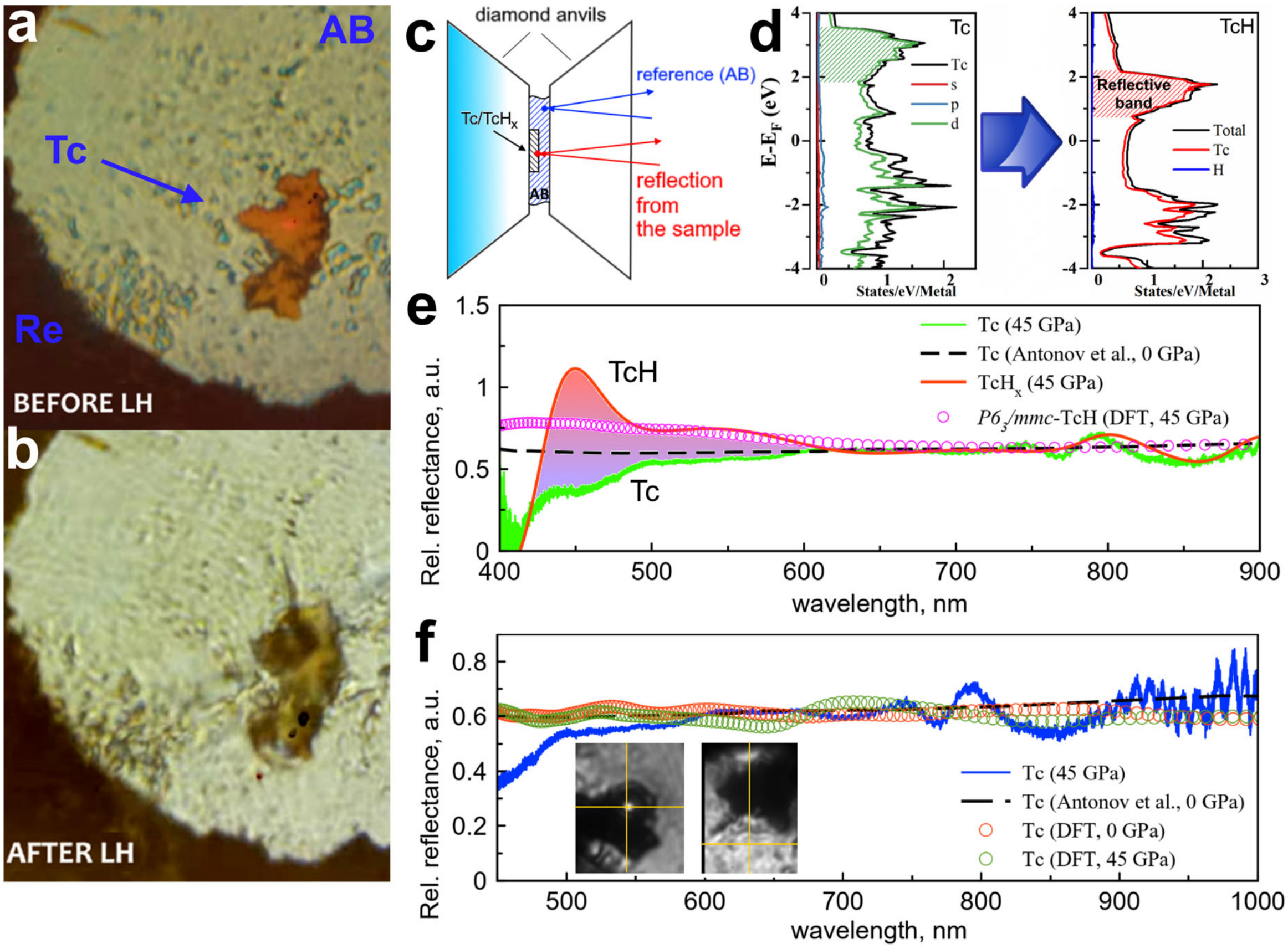

**Figure 4.** Reflectance spectroscopy of technetium (Tc) and technetium hydride (TcH) at 45 GPa in DAC T2. Microphotographs of the sample (a) before and (b) after the laser heating at 45 GPa; the pressure was practically unchanged due to the small sample volume (Re is the gasket material, AB is ammonia borane). (c) Schematic of the study of the relative reflectivity $I_{\text{sample}}/I_{\text{ref}}$. (d) Changes in the electronic density of states in the reaction $Tc + H_2 \rightarrow TcH$ at 20 GPa. (e) Smoothed measurement results of the relative reflectivity of Tc (green) and $TcH_{1+x}$ (red), calculated data for *hcp*-TcH (purple) at 45 GPa, and literature data for technetium at 0 GPa (dashed black line). (f) Experimental reflectivity of Tc at 45 GPa (blue) compared with the literature data at 0 GPa (dashed black line) and VASP calculation results at 0 (red) and 45 GPa (green).

It was found that the accuracy and reproducibility of the reflection spectra do not allow them to be used as an independent method of structure determination even in combination with the DFT calculations of reflectivity. This is primarily due to complex interference patterns and large variations of the reflection spectra in different points of non-uniform powder samples. In addition, calculations of the reflectance require the knowledge of the possible structure of a compound and are quite time-consuming compared to an almost instantaneous prediction using an XRD pattern.

Technetium hydrides are a convenient object for reflection spectroscopy studies. The analysis of the electronic structure of *hcp*-Tc and *hcp*-TcH (Figure 4d, Supporting Information Figures S6, S7) shows that the Fermi level is at a “trough” and is due to the contribution of d electrons from the Tc atoms; hydrogen makes almost no contribution to the density of states and almost does not chemically interact with Tc. When going from technetium to its hydride TcH, we see that the region of high density of unoccupied states shifts closer to the Fermi level (Figure 4d) and transitions with an energy of 2.25–2.5 eV appear near the Γ-point (Supporting Information Figure S7). This is qualitatively consistent with the results of the direct DFT calculations, which also show the emergence of a reflective band in $R(\lambda)$ at 450–600 nm. This peculiarity of the reflection spectrum is observed in the experiment with technetium hydrides but absent in the study of pure technetium before the laser heating of the sample (Figure 4e). After the laser heating of the Tc/AB sample at 45 GPa, a new reflective band (broad peak) appears at 450–600 nm, which can be explained by the presence of *hcp*-$TcH_{1+x}$. At the same time, at 500–800 nm, the relative reflectance of Tc agrees with the literature data[48] and results of the DFT calculations (Figure 4f).

## Conclusions

We investigated the formation of lower technetium hydrides at pressures up to 45 GPa using the synchrotron X-ray diffraction, reflectance spectroscopy, and first-principles calculations. In the investigated pressure range, the Tc–H system is similar to the Mo–H system: technetium reacts with hydrogen to form nonstoichiometric hydrides with gradually increasing hydrogen content up to *hcp*-$TcH_{1.3}$ at 27 GPa. The formation of technetium hydrides is also confirmed by the emergence of a reflective broad band at 450–600 nm in the hydrogenated Tc samples at 45 GPa. The dependence of the relative reflectance on the wavelength at 450–900 nm for Tc and $TcH_{1+x}$ was established. The ab initio calculations show that lower technetium hydrides do not possess superconducting properties.

Relatively simple reflectance spectroscopy provides additional evidence of hydride formation in the Tc-H system and can be used independently of synchrotron X-ray diffraction. The developed approach to the reflectance spectra studies can be useful as a complementary method to confirm the structure of polyhydrides synthesized in diamond anvil cells at ultrahigh pressures. These optical measurements do not require the use of cryogenic equipment, sputtering of electrodes, and synchrotron radiation sources, and can be implemented in almost any high-pressure laboratory.

## Author contributions

D.Z., D.V.S., and M.A.V. contributed equally to this work. D.Z., D.V.S., M.A.V., I.A.T., A.Yu.S., and D.A.S. performed the experiments. D.Z. and D.V.S. prepared the diamond anvil cells and studied the samples in them. M.A.V. and K.E.G. prepared metallic technetium for the experiments. I.A.T. carried out the laser heating of the samples in DACs. A.Yu.S. helped with the X-ray diffraction study at the Kurchatov synchrotron source, RKFM station (Moscow). D.Z. and D.V.S. prepared the theoretical analysis and calculated the equation of states, electron and phonon band structures, and superconducting and optical properties of the samples. I. V. C. performed COHP calculations using the Lobster code. D.A.S. built a system to study the optical properties of hydrides at high pressures. D.V.S., D.Z., and A.R.O. analyzed and interpreted the experimental results and wrote the manuscript. P.G.L., A.R.O., and K.E.G. directed the research, analyzed the results, and edited the manuscript. All the authors provided critical feedback and helped shape the research.

## Acknowledgments

In situ X-ray diffraction experiments at high pressures were performed on a synchrotron source of the Kurchatov Institute (KISI-Kurchatov), station RKFM. The high-pressure experiments were supported by the Ministry of Science and Higher Education of the Russian Federation within the state assignment of the FSRC Crystallography and Photonics of the RAS while radiochemistry part was carried out in frame of MSHERF topic № AAAA-A18-118040590105-4). I.A.T. was supported by the Russian Science Foundation, project No. 22-12-00163. A.R.O. thanks the Russian Science Foundation (grant 19-72-30043). D.V.S. thanks the Russian Foundation for Basic Research (project 20-32-90099) and the Russian Science Foundation, grant 22-22-00570. We also thank Igor Grishin (Skoltech) for proofreading the manuscript, and Dr. Alexander G. Kvashnin (Skoltech) for the help in calculations of dielectric functions.

## Conflict of Interest

The authors declare no conflict of interest.

## Supporting Information

The raw and processed data required to reproduce these findings are available to download from GitHub and as Supporting Information for the manuscript.

The video of the laser heating of Tc samples can be seen here: https://youtube.com/shorts/c-TePaqiEkw

The X-ray diffraction data can be found here: https://github.com/mark6871/technetium-hydrides/tree/XRD

The reflection spectroscopy data can be found here: https://github.com/mark6871/technetium-hydrides/tree/Reflection-spectroscopy

The results of the structural search are available here: https://github.com/mark6871/technetium-hydrides/tree/USPEX

# SUPPORTING INFORMATION

# Synthesis of technetium hydride $TcH_{1.3}$ at 27 GPa

D. Zhou,[1,*] D. V. Semenok,[1,*] M. A. Volkov,[2] I. A. Troyan,[3] A. Yu. Seregin,[3,4] I. V. Chepkasov,[1] D. A. Sannikov,[5] P. G. Lagoudakis,[5] A. R. Oganov,[1] and K. E. German[2,*]

[1] Skolkovo Institute of Science and Technology, Bolshoy Boulevard 30, bld. 1, Moscow 121205, Russia
[2] A. N. Frumkin Institute of Physical Chemistry and Electrochemistry of the Russian Academy of Sciences, Department of Radiochemistry, 31 Leninsky Prospekt, Moscow 119991, Russia
[3] Shubnikov Institute of Crystallography, Federal Scientific Research Center Crystallography and Photonics, Russian Academy of Sciences, 59 Leninsky Prospekt, Moscow 119333, Russia
[4] National Research Center "Kurchatov Institute", Ploshchad' Akademika Kurchatova, 1, Moscow 123182, Russia
[5] Center for Photonics and Quantum Materials, Skolkovo Institute of Science and Technology, Bolshoy Boulevard 30, bld. 1, Moscow 121205, Russia

*Corresponding authors: d.zhou@skoltech.ru, dmitrii.semenok@skoltech.ru, guerman_k@mail.ru

## Contents

## Methods

*Metallic technetium*

Metallic technetium was obtained by thermal decomposition of an ammonium pertechnetate in a stream of hydrogen. A sample of ammonium pertechnetate was placed in an $Al_2O_3$ crucible and kept in a furnace at 350 °C for 15 minutes. Then the temperature was raised to 900 °C and the sample was incubated in a hydrogen atmosphere during 1 hour for the complete reduction of formed $TcO_2$ to metal. The sample was cooled to room temperature, also in a stream of hydrogen. The ammonium pertechnetate (from the "Isotope" Ltd.) used in this work was previously purified by recrystallization from double distilled water. The resulting white crystalline powder was characterized using the powder X-ray diffraction on a Panalytical AERIS diffractometer. No additional phases were detected in the Tc powder under study. The powder X-ray diffraction of the sample with a metallic luster showed only peaks corresponding to metallic technetium with a hexagonal dense packed crystal lattice.

*Caution:* technetium-99 is a weak beta emitter ($E_{max}$ = 292 keV, specific activity 635 Bq/micro-g). All manipulations with this nuclide were performed in a radiochemistry laboratory of IPCE RAS designed for chemical synthesis with radionuclides using efficient HEPA-filtered fume hoods, Schlenk and glove box techniques, and following locally approved radioisotope handling and monitoring procedures.

*Equation of states*

The calculations of the equation of state of technetium in the VASP code[1–3] using the PAW potentials derived within the PBE functional, pseudopotentials with 13 valence electrons ($4p^6 5s^2 4d^5$) lead to an error in the unit cell volume of *hcp*-Tc of about 2% per each Tc atom (Figure S3). To take this error into account, the VASP calculations of the equations of state for various $Tc_yH_x$ hydrides were corrected using the formula $V = V(Tc_yH_x) - 0.02 \times y \times V(Tc)$.

*XRD and Raman measurements*

The XRD patterns of the $TcH_x$ samples at room temperature were recorded at the RKFM beamline of the Kurchatov Institute (KISI-Kurchatov, Moscow) using monochromatic synchrotron radiation and MAR165 detector. The X-ray beam with a wavelength of 0.62 Å was focused in a 50 × 50 μm spot. The XRD data were analyzed and integrated using Dioptas software package (version 0.5.3).[4] The full profile analysis of the diffraction patterns and the calculation of the unit cell parameters were performed using JANA2006 software[5] with the Le Bail method.[6] The pressure in diamond anvil cells (DACs) was determined via the Raman signal of diamond at room temperature[7] using LabRAM HR Evolution (Horiba) and DXR3xi Raman Imaging Microscope (Thermo Scientific). In both cases, 532 nm excitation laser light was used. To synthesize technetium hydrides, we used diamond anvils with culet diameters of 200 and 300 μm, rhenium gaskets, symmetrical Mao-type DACs, and ammonia borane $NH_3BH_3$ (AB) as a hydrogen source. Laser heating was carried out by 2-3 second pulses of IR laser (~1 μm), which led to the local temperature rise above 1000 K. The sizes of loaded technetium particles ranged from 20 to 60 microns.

*Structural search and superconductivity*

The computational predictions of thermodynamically stable Tc–H phases at 5, 25, and 50 GPa were carried out using the variable-composition evolutionary algorithm USPEX.[8–10] The evolutionary searches were combined with structure relaxations using the density functional theory (DFT)[11,12] within the Perdew–Burke–Ernzerhof (PBE) generalized gradient approximation (GGA) functional[13] and the projector augmented wave method[14,15] as implemented in the VASP and Quantum Espresso codes.[16,17] The kinetic energy cutoff for plane waves was 600 eV. The Brillouin zone was sampled using Γ-centered *k*-points meshes with a resolution of $2\pi \times 0.05$ Å$^{-1}$. We also calculated the phonon densities of states of the studied materials using the finite displacements method (VASP and PHONOPY[18,19]).

The calculations of the critical temperature of superconductivity $T_C$ were carried out using Quantum ESPRESSO (QE) package.[16,17] The phonon frequencies and electron–phonon coupling (EPC) coefficients were computed using the density functional perturbation theory,[20] employing the plane-wave pseudopotential method and the PBE exchange–correlation functional. Within the optimized tetrahedron method,[21] we calculated the electron–phonon coupling coefficients λ and the Eliashberg functions via sampling of the first Brillouin zone by 12×12×8 *k*-points and 3×3×2 *q*-points meshes. To evaluate the density of states (DOS) and phonon linewidths, a denser 16×16×12 *k*-mesh was used. The critical temperature $T_C$ was calculated using the Allen–Dynes formula.[22]

*Optical properties*

The calculation of optical properties was performed using the VASP code according to the methodology of Gajdoš et al. [23] (with LOPTICS = .TRUE.), the number of frequency grid points was 20000 (NEDOS), the total number of Kohn-Sham orbitals in the calculation was 120 (NBANDS), cutoff energy was 600 eV. Local field effects were included on the Hartree level only (LRPA=.TRUE., CSHIFT = 0.1).

## Crystal structures

| **_hcp_-Tc (0 GPa)** | **_hcp_-TcH (50 GPa)** |
|---|---|
| # CIF file<br>data_findsym-output<br><br>_symmetry_space_group_name_H-M 'P 63/m 2/m 2/c'<br>_symmetry_Int_Tables_number 194<br>_cell_length_a 2.76100<br>_cell_length_b 2.76100<br>_cell_length_c 8.83700<br>_cell_angle_alpha 90.00000<br>_cell_angle_beta 90.00000<br>_cell_angle_gamma 120.00000<br>_cell_volume 58.3402<br><br>loop_<br>_space_group_symop_operation_xyz<br>x,y,z<br>x-y,x,z+1/2<br>-y,x-y,z<br>-x,-y,z+1/2<br>-x+y,-x,z<br>y,-x+y,z+1/2<br>x-y,-y,-z<br>x,x-y,-z+1/2<br>y,x,-z<br>-x+y,y,-z+1/2<br>-x,-x+y,-z<br>-y,-x,-z+1/2<br>-x,-y,-z<br>-x+y,-x,-z+1/2<br>y,-x+y,-z<br>x,y,-z+1/2<br>x-y,x,-z<br>-y,x-y,-z+1/2<br>-x+y,y,z<br>-x,-x+y,z+1/2<br>-y,-x,z<br>x-y,-y,z+1/2<br>x,x-y,z<br>y,x,z+1/2<br><br>loop_<br>_atom_site_label<br>_atom_site_type_symbol<br>_atom_site_fract_x<br>_atom_site_fract_y<br>_atom_site_fract_z<br>_atom_site_occupancy<br>Tc1 Tc 0.00000 0.00000 0.00000 1.00000<br>Tc2 Tc 0.33333 0.66667 0.25000 1.00000<br><br># end_of_file | # CIF file<br>data_findsym-output<br><br>_symmetry_space_group_name_H-M 'P 63/m 2/m 2/c'<br>_symmetry_Int_Tables_number 194<br>_cell_length_a 2.77802<br>_cell_length_b 2.77802<br>_cell_length_c 4.31200<br>_cell_angle_alpha 90.00000<br>_cell_angle_beta 90.00000<br>_cell_angle_gamma 120.00000<br><br>loop_<br>_space_group_symop_operation_xyz<br>x,y,z<br>x-y,x,z+1/2<br>-y,x-y,z<br>-x,-y,z+1/2<br>-x+y,-x,z<br>y,-x+y,z+1/2<br>x-y,-y,-z<br>x,x-y,-z+1/2<br>y,x,-z<br>-x+y,y,-z+1/2<br>-x,-x+y,-z<br>-y,-x,-z+1/2<br>-x,-y,-z<br>-x+y,-x,-z+1/2<br>y,-x+y,-z<br>x,y,-z+1/2<br>x-y,x,-z<br>-y,x-y,-z+1/2<br>-x+y,y,z<br>-x,-x+y,z+1/2<br>-y,-x,z<br>x-y,-y,z+1/2<br>x,x-y,z<br>y,x,z+1/2<br><br>loop_<br>_atom_site_label<br>_atom_site_type_symbol<br>_atom_site_fract_x<br>_atom_site_fract_y<br>_atom_site_fract_z<br>_atom_site_occupancy<br>Tc1 Tc 0.33333 0.66667 0.25000 1.00000<br>H1 H 0.00000 0.00000 0.00000 1.00000<br><br># end_of_file |

**$P1$-$Tc_{12}H_9$ (pseudohexagonal, 50 GPa)**

```
# CIF file

data_findsym-output
 _symmetry_space_group_name_H-M 'P 1'
_symmetry_Int_Tables_number 1

_cell_length_a       6.41800
_cell_length_b       6.41700
_cell_length_c       4.74000
_cell_angle_alpha  111.66200
_cell_angle_beta    68.35100
_cell_angle_gamma  100.41700

loop_
_space_group_symop_operation_xyz
x,y,z

loop_
_atom_site_label
_atom_site_type_symbol
_atom_site_fract_x
_atom_site_fract_y
_atom_site_fract_z
_atom_site_occupancy
Tc1 Tc  -0.31412   0.04078   0.09887   1.00000
Tc2 Tc  -0.47746  -0.12145  -0.40104   1.00000
Tc3 Tc   0.19241  -0.46540  -0.40452   1.00000
Tc4 Tc   0.02753   0.36916   0.09015   1.00000
Tc5 Tc   0.27851   0.12216   0.18441   1.00000
Tc6 Tc   0.11326  -0.03920  -0.32546   1.00000
Tc7 Tc  -0.14190   0.20527  -0.40042   1.00000
Tc8 Tc  -0.05768  -0.20980   0.18143   1.00000
Tc9 Tc  -0.22508  -0.37726  -0.31890   1.00000
Tc10 Tc  -0.39162   0.45188   0.17695   1.00000
Tc11 Tc   0.43915   0.28216  -0.32372   1.00000
Tc12 Tc   0.35866  -0.29179   0.09868   1.00000
H1  H    0.06764  -0.33683   0.38968   1.00000
H2  H    0.23221  -0.16865  -0.11371   1.00000
H3  H    0.48394  -0.41944  -0.36228   1.00000
H4  H    0.32381   0.40672   0.12985   1.00000
H5  H   -0.26590   0.33027   0.39014   1.00000
H6  H   -0.35517  -0.24719   0.14540   1.00000
H7  H   -0.01629   0.08010   0.13715   1.00000
H8  H   -0.18984  -0.07995  -0.35424   1.00000
H9  H   -0.09486   0.49306  -0.11686   1.00000

# end_of_file
```

**$P$-1-$Tc_{12}H_7$ (pseudohexagonal, 50 GPa)**

```
# CIF file

data_findsym-output
 _symmetry_space_group_name_H-M 'P -1'
_symmetry_Int_Tables_number 2

_cell_length_a       6.93000
_cell_length_b       5.45000
_cell_length_c       5.08600
_cell_angle_alpha  105.46000
_cell_angle_beta    96.18900
_cell_angle_gamma  113.03700

loop_
_space_group_symop_operation_xyz
x,y,z
-x,-y,-z

loop_
_atom_site_label
_atom_site_type_symbol
_atom_site_fract_x
_atom_site_fract_y
_atom_site_fract_z
_atom_site_occupancy
Tc1 Tc  -0.02372   0.16654  -0.28519   1.00000
Tc2 Tc   0.35757   0.33474  -0.37934   1.00000
Tc3 Tc  -0.30623  -0.16947  -0.05053   1.00000
Tc4 Tc  -0.35635   0.16936   0.38732   1.00000
Tc5 Tc   0.02358   0.33256   0.28481   1.00000
Tc6 Tc   0.30567  -0.33785   0.04917   1.00000
H1  H    0.33264   0.49895   0.33568   1.00000
H2  H    0.00000   0.00000   0.00000   1.00000
H3  H    0.50000   0.00000   0.00000   1.00000
H4  H   -0.32992  -0.00163  -0.33994   1.00000
H5  H    0.00000   0.50000   0.00000   1.00000

# end_of_file
```

### *P*-1-$Tc_8H_7$ (pseudohexagonal, 50 GPa)

```
#CIF file
data_findsym-output

_symmetry_space_group_name_H-M 'P -1'
_symmetry_Int_Tables_number 2

_cell_length_a       6.43800
_cell_length_b       5.50600
_cell_length_c       6.43900
_cell_angle_alpha   89.99100
_cell_angle_beta    84.89700
_cell_angle_gamma   90.00300

loop_
_space_group_symop_operation_xyz
x,y,z
-x,-y,-z

loop_
_atom_site_label
_atom_site_type_symbol
_atom_site_fract_x
_atom_site_fract_y
_atom_site_fract_z
_atom_site_occupancy
Tc1 Tc  -0.16115   0.37810  -0.41966   1.00000
Tc2 Tc  -0.16527  -0.12788  -0.41867   1.00000
Tc3 Tc   0.08470   0.12788  -0.16905   1.00000
Tc4 Tc   0.08464  -0.37797  -0.16897   1.00000
Tc5 Tc   0.41715  -0.12542  -0.33454   1.00000
Tc6 Tc   0.41334   0.37549  -0.33374   1.00000
Tc7 Tc   0.33246  -0.12550   0.08417   1.00000
Tc8 Tc   0.33242   0.37541   0.08402   1.00000
H1  H   -0.12534   0.12478   0.37470   1.00000
H2  H   -0.11955   0.37361  -0.13106   1.00000
H3  H    0.36921   0.37823   0.38067   1.00000
H4  H    0.12214   0.12350   0.12502   1.00000
H5  H    0.37561  -0.37473  -0.12483   1.00000
H6  H    0.37212  -0.12847   0.37497   1.00000
H7  H    0.37553   0.12475  -0.12474   1.00000

# end_of_file
```

### *R*-3*m*-TcH (5 GPa)

```
#CIF file
data_findsym-output
 _symmetry_space_group_name_H-M 'R -3 2/m (hexagonal axes)'
_symmetry_Int_Tables_number 166
_cell_length_a       2.85558
_cell_length_b       2.85558
_cell_length_c      20.55822
_cell_angle_alpha   90.00000
_cell_angle_beta    90.00000
_cell_angle_gamma  120.00000

loop_
_space_group_symop_operation_xyz
x,y,z
-y,x-y,z
-x+y,-x,z
y,x,-z
-x,-x+y,-z
x-y,-y,-z
-x,-y,-z
y,-x+y,-z
x-y,x,-z
-y,-x,z
x,x-y,z
-x+y,y,z
x+1/3,y+2/3,z+2/3
-y+1/3,x-y+2/3,z+2/3
-x+y+1/3,-x+2/3,z+2/3
y+1/3,x+2/3,-z+2/3
-x+1/3,-x+y+2/3,-z+2/3
x-y+1/3,-y+2/3,-z+2/3
-x+1/3,-y+2/3,-z+2/3
y+1/3,-x+y+2/3,-z+2/3
x-y+1/3,x+2/3,-z+2/3
-y+1/3,-x+2/3,z+2/3
x+1/3,x-y+2/3,z+2/3
-x+y+1/3,y+2/3,z+2/3
x+2/3,y+1/3,z+1/3
-y+2/3,x-y+1/3,z+1/3
-x+y+2/3,-x+1/3,z+1/3
y+2/3,x+1/3,-z+1/3
-x+2/3,-x+y+1/3,-z+1/3
x-y+2/3,-y+1/3,-z+1/3
-x+2/3,-y+1/3,-z+1/3
y+2/3,-x+y+1/3,-z+1/3
x-y+2/3,x+1/3,-z+1/3
-y+2/3,-x+1/3,z+1/3
x+2/3,x-y+1/3,z+1/3
-x+y+2/3,y+1/3,z+1/3

loop_
_atom_site_label
_atom_site_type_symbol
_atom_site_fract_x
_atom_site_fract_y
_atom_site_fract_z
_atom_site_occupancy
Tc1 Tc   0.00000   0.00000   0.27983   1.00000
Tc2 Tc   0.00000   0.00000   0.50000   1.00000
H1  H    0.00000   0.00000   0.11441   1.00000
H2  H    0.00000   0.00000   0.00000   1.00000

# end_of_file
```

| ***P2/m*-$Tc_4H_3$ (pseudohexagonal, 5 GPa)** | **C2/*m*-$Tc_8H_5$ (pseudohexagonal, 5 GPa)** |
|---|---|
| # CIF file<br>data_findsym-output<br><br>_symmetry_space_group_name_H-M 'P 1 2/m 1'<br>_symmetry_Int_Tables_number 10<br><br>_cell_length_a 4.47200<br>_cell_length_b 2.84600<br>_cell_length_c 4.92800<br>_cell_angle_alpha 90.00000<br>_cell_angle_beta 90.03800<br>_cell_angle_gamma 90.00000<br><br>loop_<br>_space_group_symop_operation_xyz<br>x,y,z<br>-x,y,-z<br>-x,-y,-z<br>x,-y,z<br><br>loop_<br>_atom_site_label<br>_atom_site_type_symbol<br>_atom_site_fract_x<br>_atom_site_fract_y<br>_atom_site_fract_z<br>_atom_site_occupancy<br>Tc1 Tc -0.25704 0.00000 0.33934 1.00000<br>Tc2 Tc 0.25245 0.50000 0.17289 1.00000<br>H1 H 0.50000 0.00000 0.00000 1.00000<br>H2 H 0.00000 0.50000 0.50000 1.00000<br>H3 H 0.00000 0.00000 0.00000 1.00000<br><br># end_of_file | # CIF file<br>data_findsym-output<br><br>_symmetry_space_group_name_H-M 'C 1 2/m 1'<br>_symmetry_Int_Tables_number 12<br><br>_cell_length_a 9.79716<br>_cell_length_b 5.64700<br>_cell_length_c 4.46900<br>_cell_angle_alpha 90.00000<br>_cell_angle_beta 89.96480<br>_cell_angle_gamma 90.00000<br><br>loop_<br>_space_group_symop_operation_xyz<br>x,y,z<br>-x,y,-z<br>-x,-y,-z<br>x,-y,z<br>x+1/2,y+1/2,z<br>-x+1/2,y+1/2,-z<br>-x+1/2,-y+1/2,-z<br>x+1/2,-y+1/2,z<br><br>loop_<br>_atom_site_label<br>_atom_site_type_symbol<br>_atom_site_fract_x<br>_atom_site_fract_y<br>_atom_site_fract_z<br>_atom_site_occupancy<br>Tc1 Tc -0.41545 0.25325 -0.24983 1.00000<br>Tc2 Tc -0.33029 0.00000 0.24230 1.00000<br>Tc3 Tc 0.16706 0.00000 0.24526 1.00000<br>H1 H 0.00000 0.50000 0.50000 1.00000<br>H2 H 0.00000 0.00000 0.00000 1.00000<br>H3 H 0.00000 0.50000 0.00000 1.00000<br>H4 H 0.25000 0.25000 0.50000 1.00000<br><br># end_of_file |

***Pm*-$Tc_{16}H_7$ (pseudohexagonal, 5 GPa)**

```
# CIF file
data_findsym-output

_symmetry_space_group_name_H-M 'P 1 m 1'
_symmetry_Int_Tables_number 6

_cell_length_a      10.14300
_cell_length_b       2.80900
_cell_length_c       9.68600
_cell_angle_alpha   90.00000
_cell_angle_beta    61.51800
_cell_angle_gamma   90.00000

loop_
_space_group_symop_operation_xyz
x,y,z
x,-y,z

loop_
_atom_site_label
_atom_site_type_symbol
_atom_site_fract_x
_atom_site_fract_y
_atom_site_fract_z
_atom_site_occupancy
Tc1 Tc   0.17328   0.00000   0.36466   1.00000
Tc2 Tc  -0.33448   0.00000   0.12122   1.00000
Tc3 Tc   0.17375   0.50000  -0.38038   1.00000
Tc4 Tc  -0.07920   0.50000  -0.42514   1.00000
Tc5 Tc   0.42433   0.00000  -0.41888   1.00000
Tc6 Tc  -0.33523   0.50000  -0.12496   1.00000
Tc7 Tc  -0.32914   0.50000   0.37668   1.00000
Tc8 Tc   0.42964   0.50000  -0.17237   1.00000
Tc9 Tc  -0.07396   0.00000  -0.17054   1.00000
Tc10 Tc   0.16539   0.50000   0.12714   1.00000
Tc11 Tc  -0.07085   0.50000   0.07663   1.00000
Tc12 Tc   0.16513   0.00000  -0.12453   1.00000
Tc13 Tc   0.41742   0.50000   0.32723   1.00000
Tc14 Tc  -0.33047   0.00000  -0.38165   1.00000
Tc15 Tc   0.42370   0.00000   0.08176   1.00000
Tc16 Tc  -0.07555   0.00000   0.33305   1.00000
H1  H   -0.46275   0.00000   0.35538   1.00000
H2  H   -0.19619   0.00000   0.22355   1.00000
H3  H    0.05384   0.00000  -0.40669   1.00000
H4  H    0.29600   0.00000  -0.02276   1.00000
H5  H    0.29571   0.50000  -0.27325   1.00000
H6  H   -0.20325   0.50000  -0.02445   1.00000
H7  H   -0.21140   0.00000  -0.27170   1.00000

# end_of_file
```

## Convex hulls

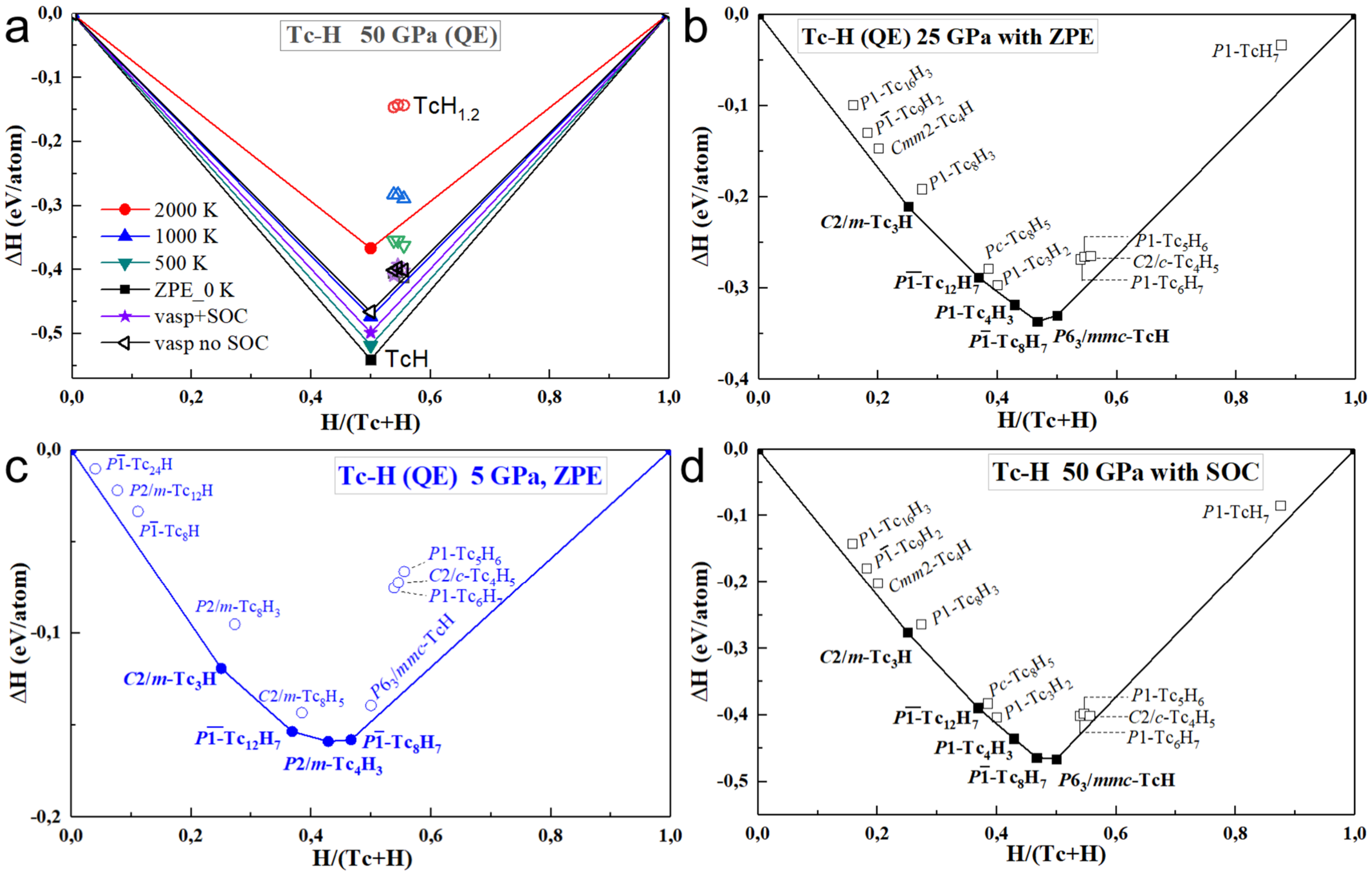

**Figure S1.** Thermodynamic stability diagrams of various pseudohexagonal phases of technetium hydrides $TcH_{1-x}$: (a) at different temperatures (0–2000 K) and 50 GPa, (b) at 25 GPa with taking into account the zero-point energy (ZPE), (c) at 5 GPa with ZPE, (d) at 50 GPa with taking into account the spin-orbit coupling (SOC). None of the additional effects leads to the stabilization of hydrides with stoichiometry above 1:1 (TcH). On the contrary, high temperatures destabilize higher technetium hydrides.

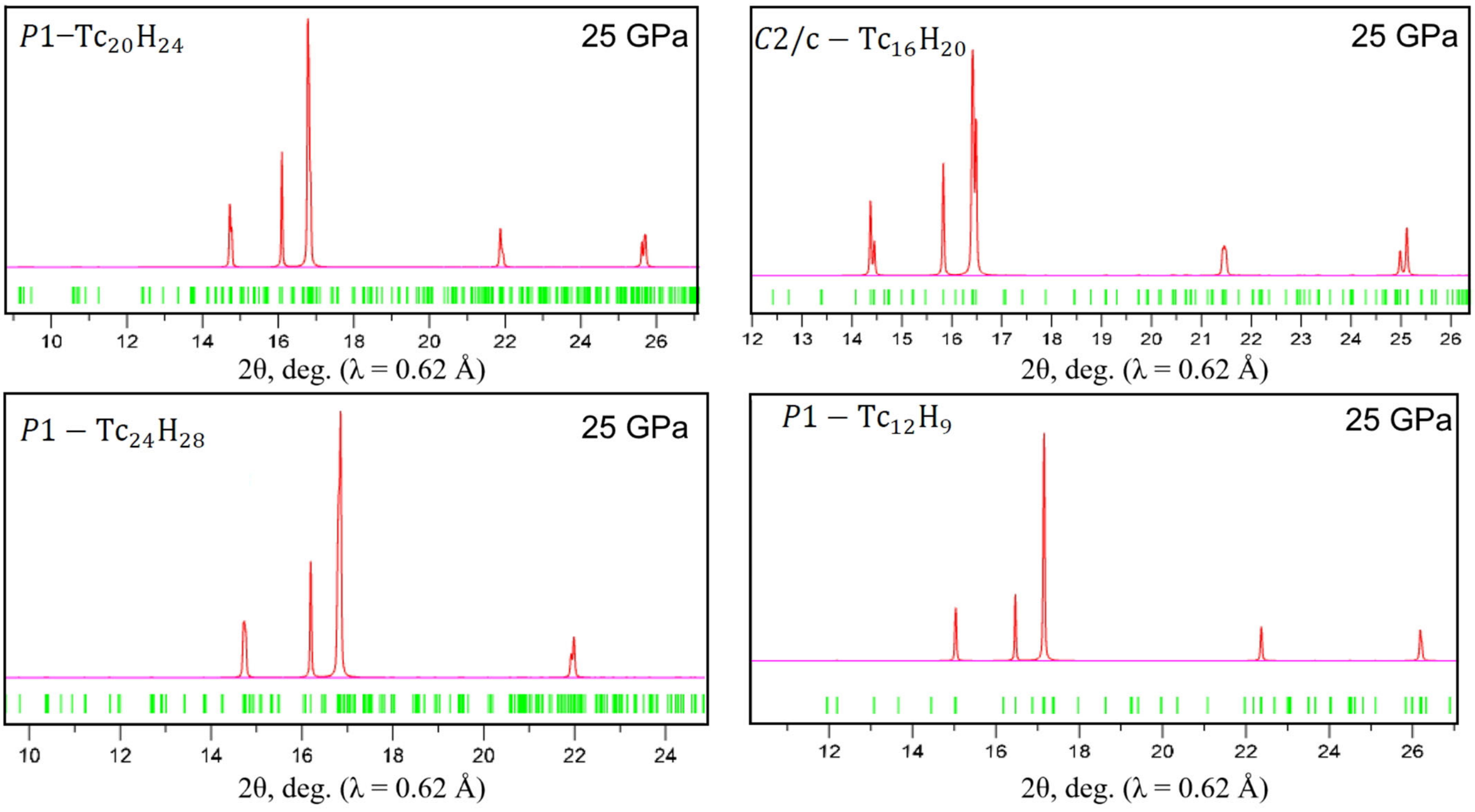

**Figure S2.** Calculated diffraction patterns (VESTA[24]) for some technetium hydrides at 25 GPa. All these compounds are pseudohexagonal and cannot be distinguished by the X-ray diffraction.

## Theoretical calculations

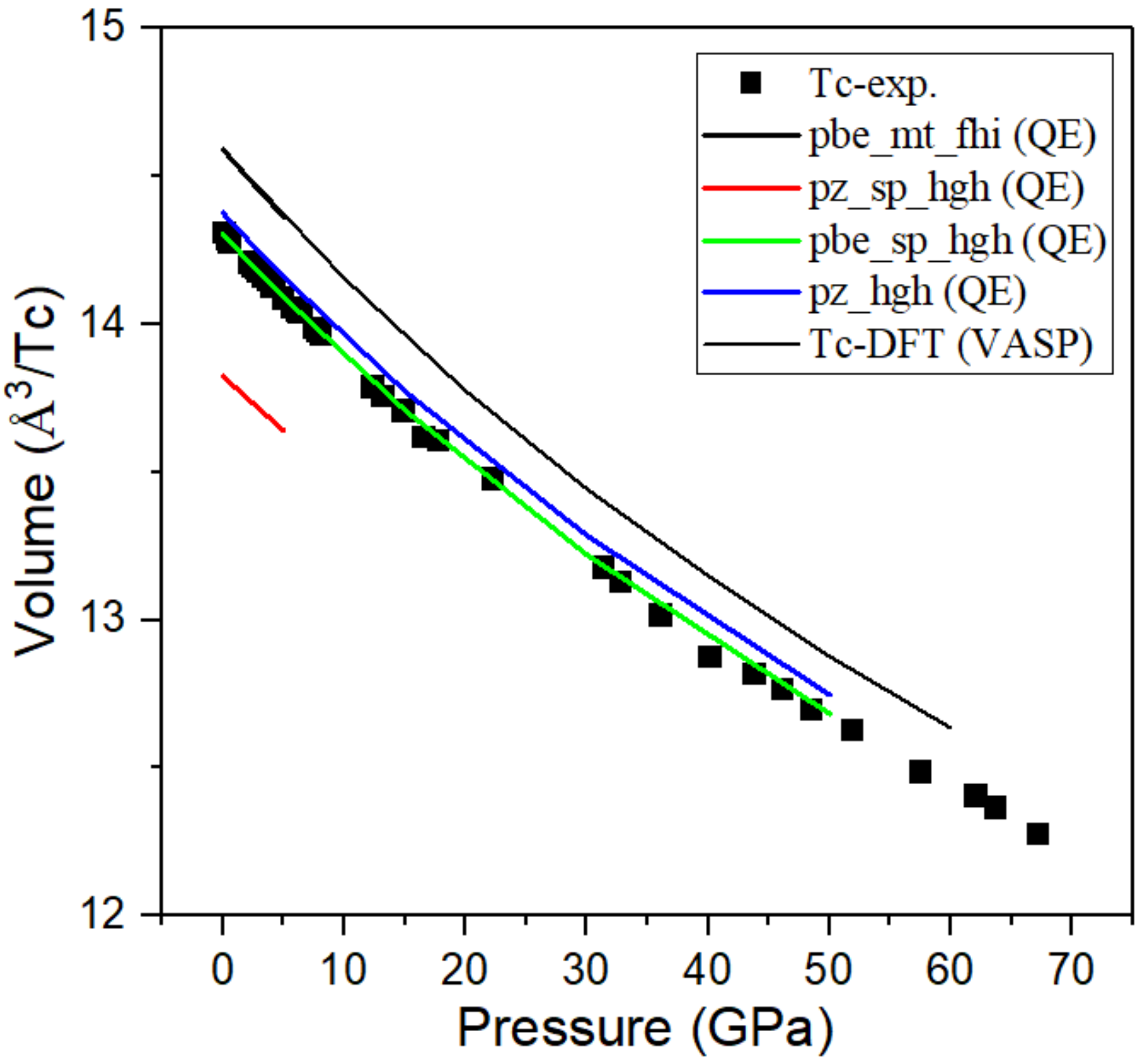

**Figure S3**. Equations of state of metallic *hcp* technetium calculated using different pseudopotentials in Quantum Espresso (QE)[16,17] and VASP[1–3] compared with the experimental results (black squares). Quantum Espresso (PBE–SP–HGH) gives a good approximation to the experimental data.

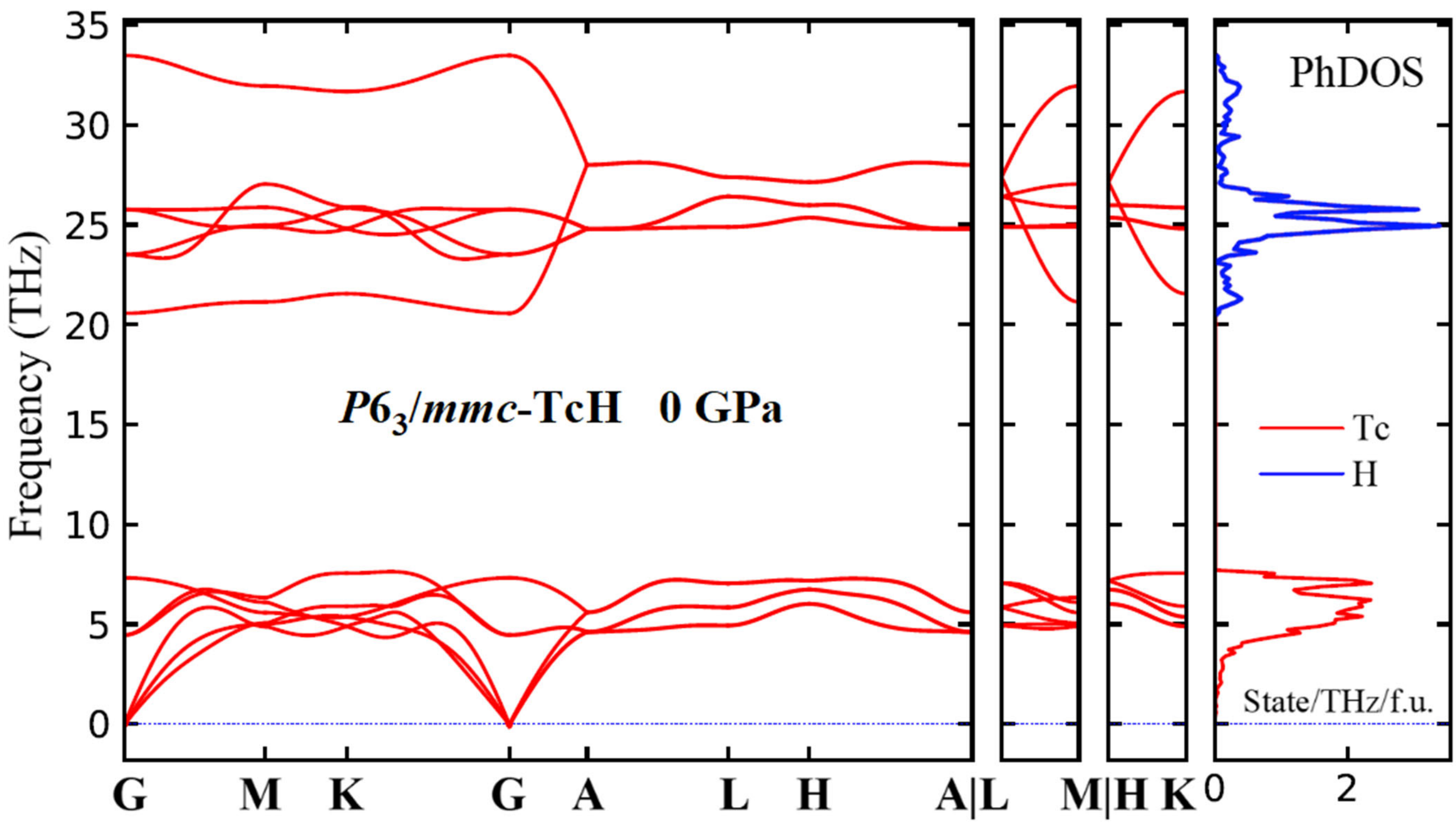

**Figure S4.** Phonon band structure and density of states of technetium monohydride at 0 GPa. The compound is dynamically stable. The results coincide with those of Ref. [25].

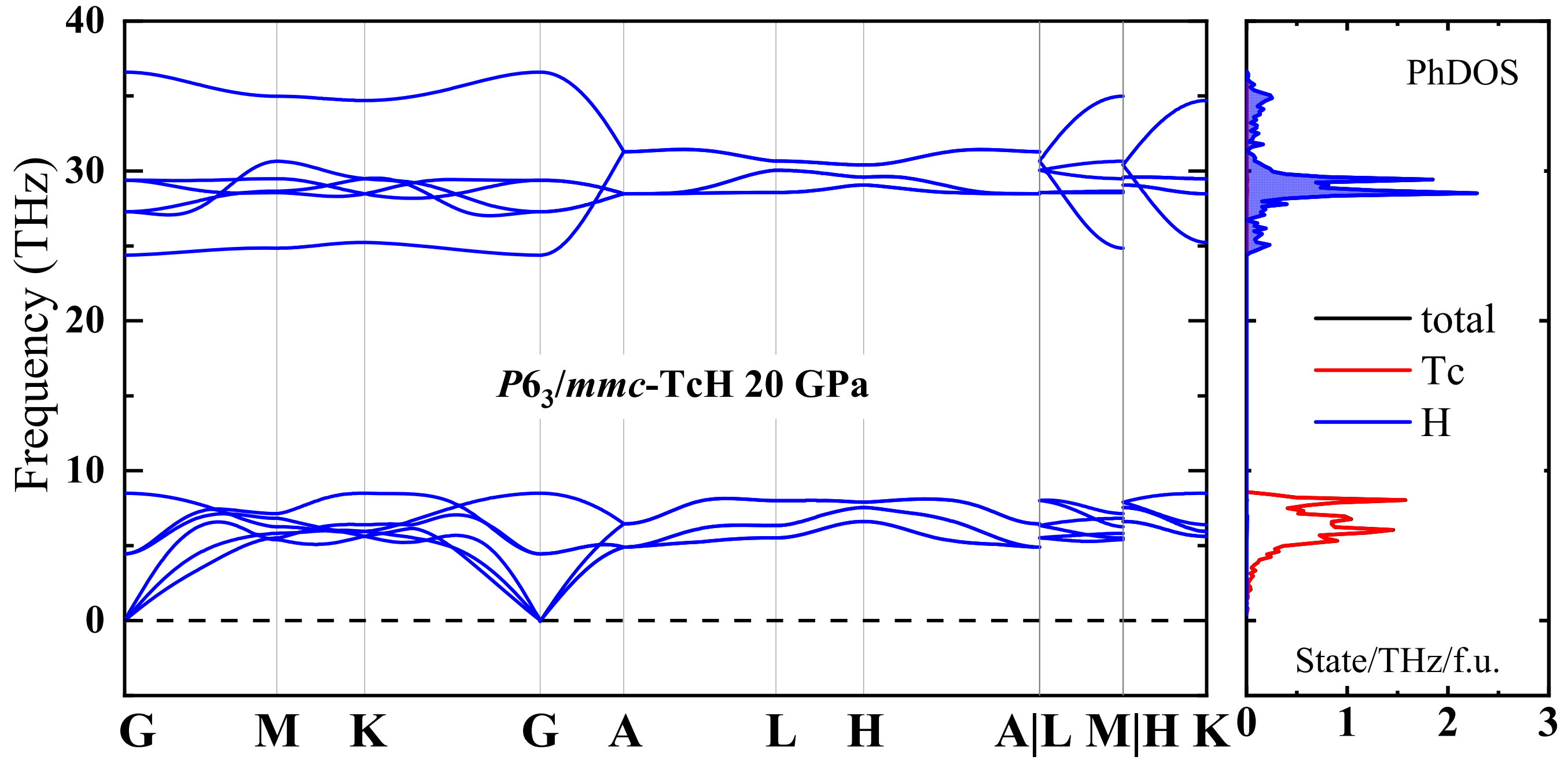

**Figure S5.** Phonon band structure and density of states of technetium monohydride at 20GPa. The compound is dynamically stable. The results coincide with those of Ref. [25].

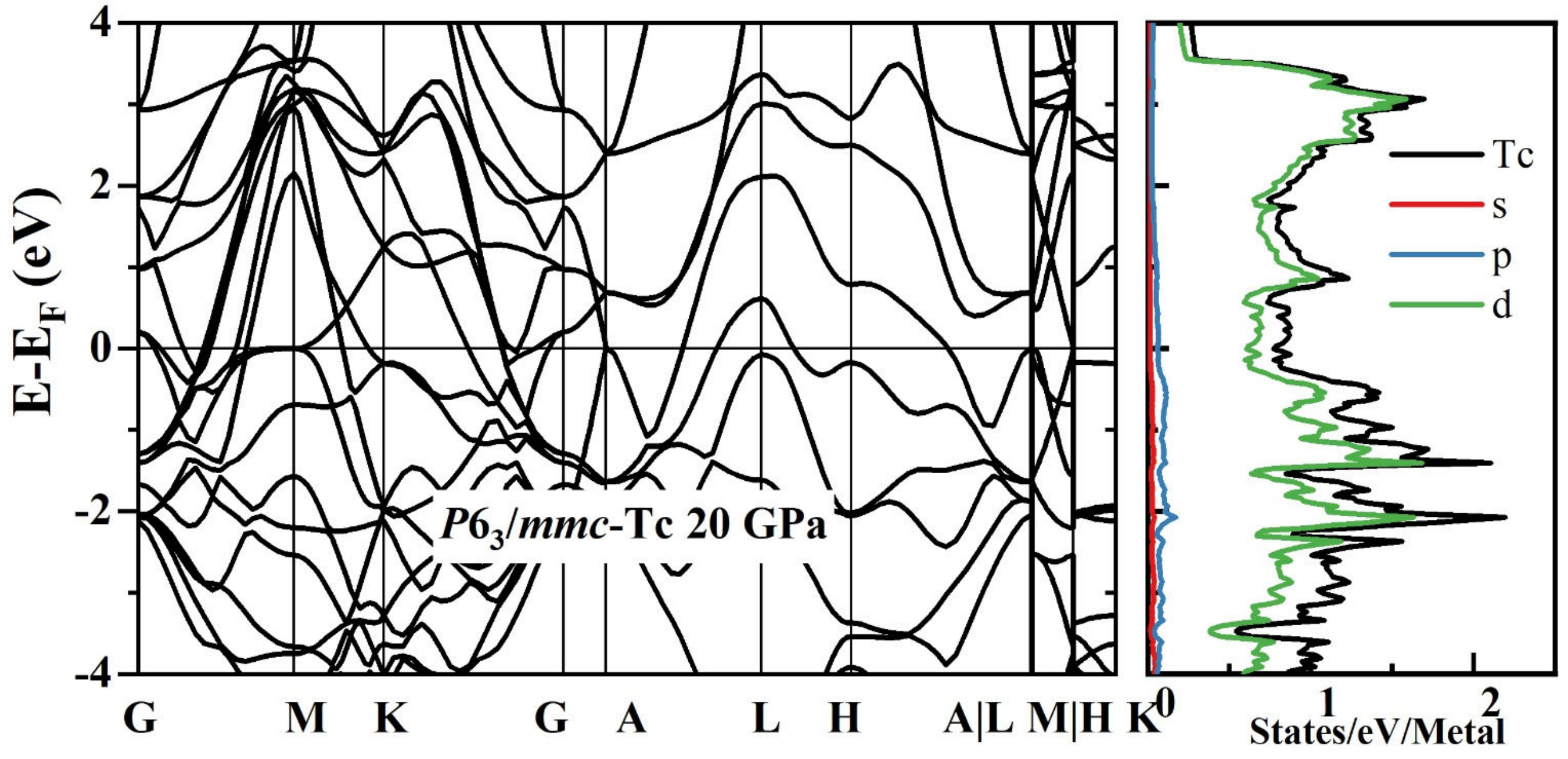

**Figure S6**. Electron band structure and density of states of technetium at 20 GPa.

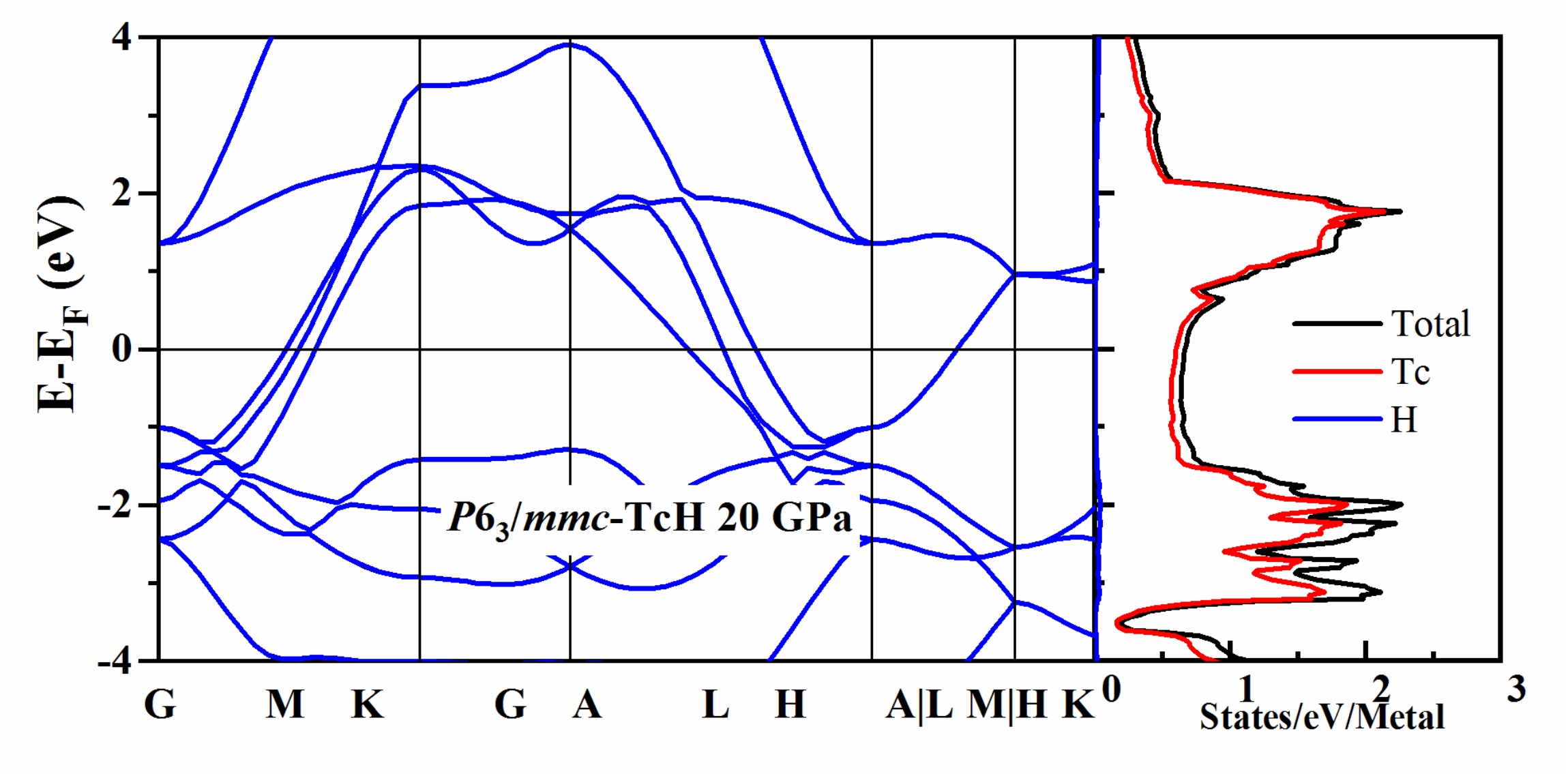

**Figure S7**. Electron band structure and density of states of technetium monohydride TcH at 20 GPa.

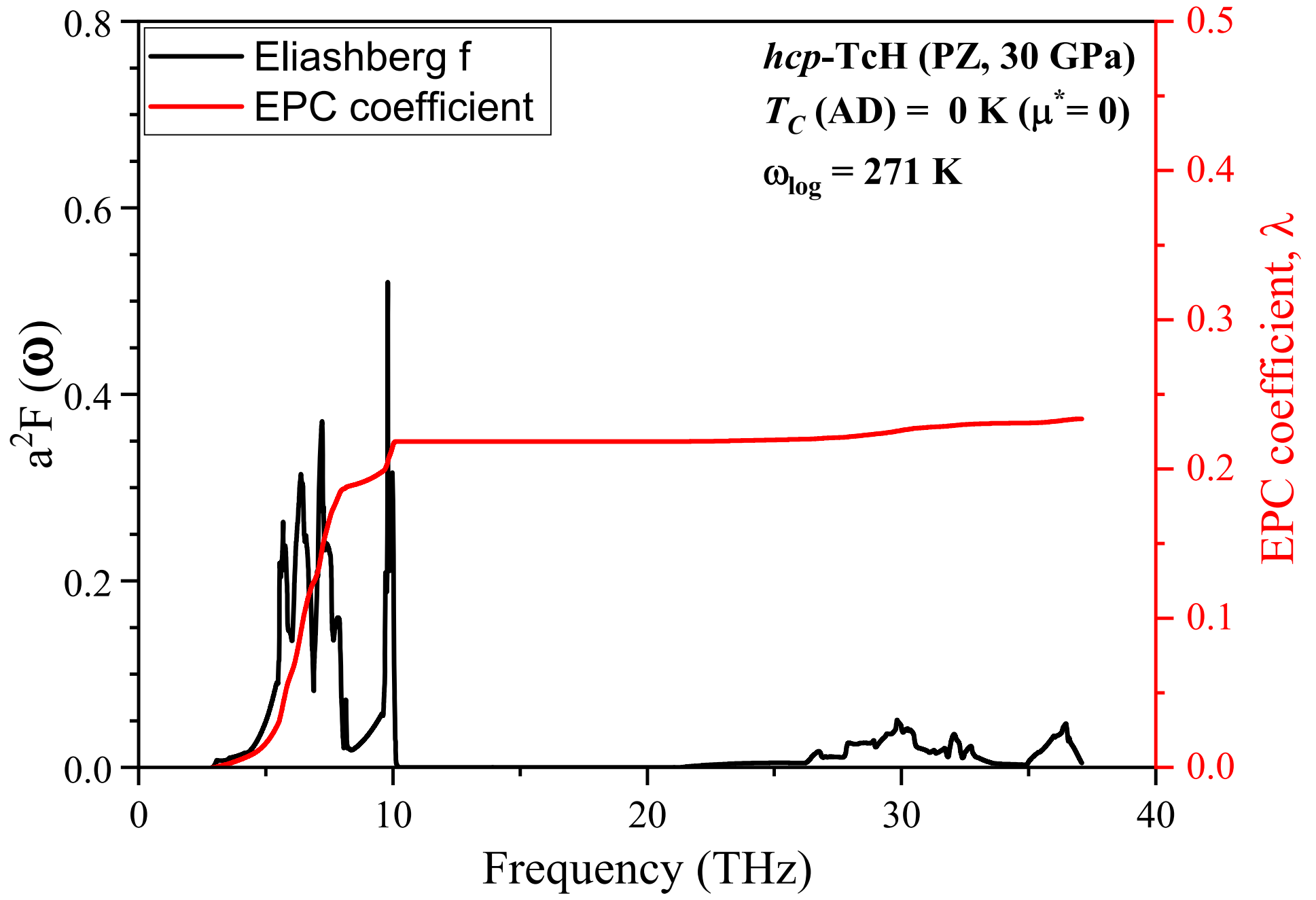

**Figure S8**. Eliashberg function and electron–phonon interaction parameter (λ) for *hcp* technetium monohydride at 30 GPa calculated using the Quantum Espresso code with the HGH-type pseudopotential and Perdew–Zunger (PZ) functional.[26,27]

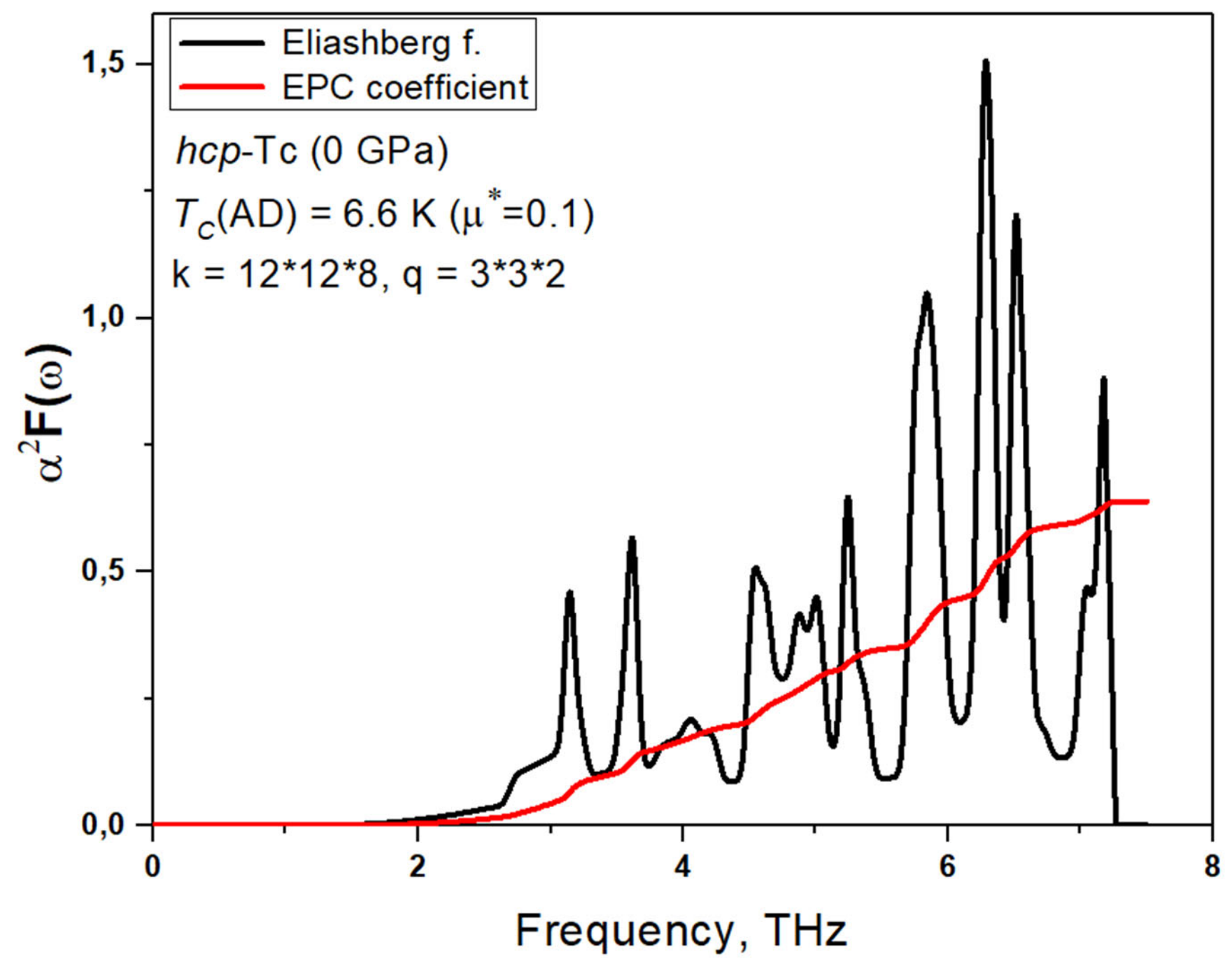

**Figure S9**. Eliashberg function and electron–phonon interaction parameter(λ) for *hcp*-Tc technetium at 0 GPa calculated using the Quantum Espresso code with the PBE functional and HGH-type pseudopotential.[26,27]

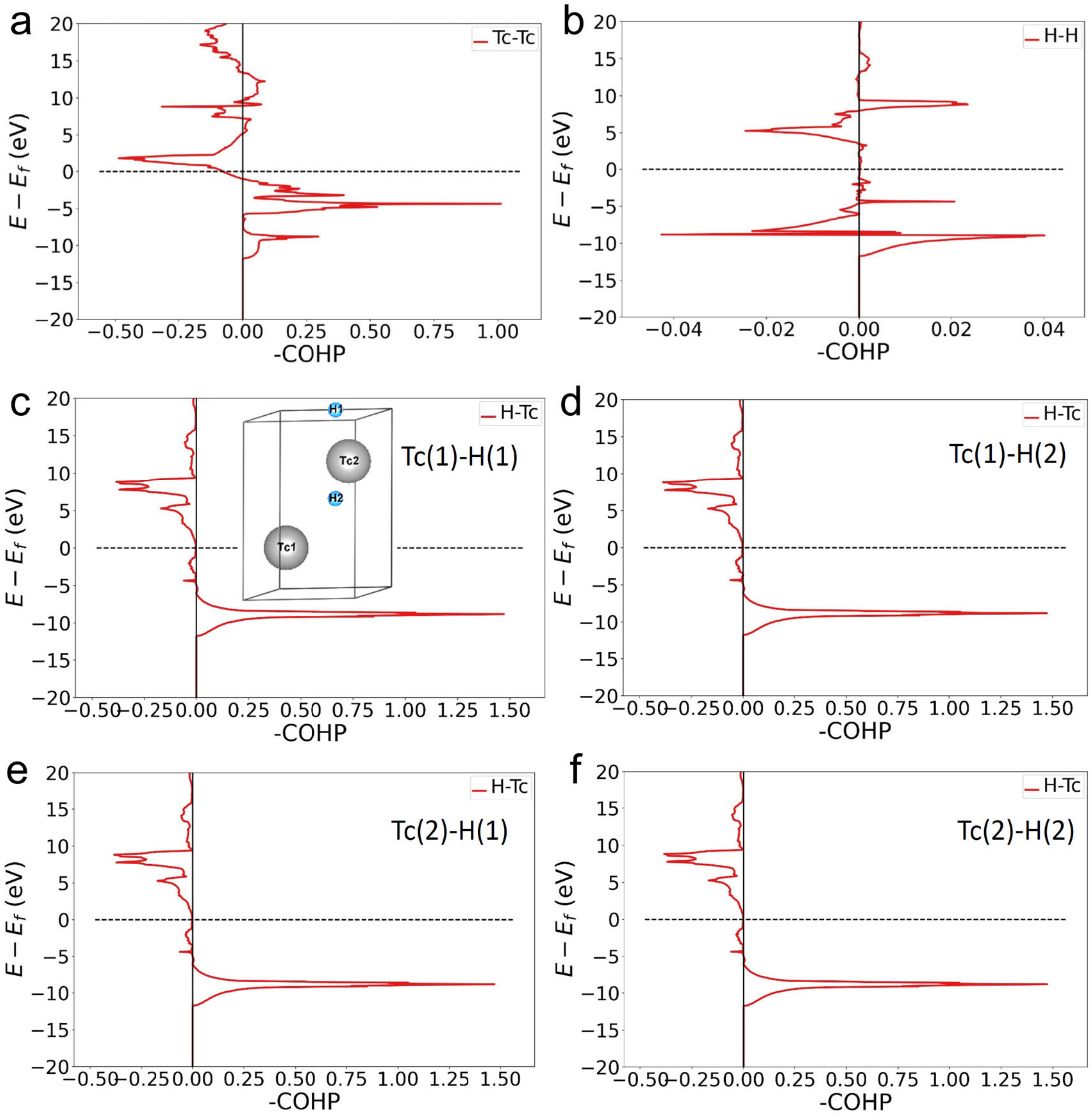

**Figure S10.** The Crystal Orbital Hamilton Population (COHP) analysis of bonding in *hcp*-TcH at 25 GPa. The calculations were performed using the Lobster code. [28]

## Raman spectroscopy

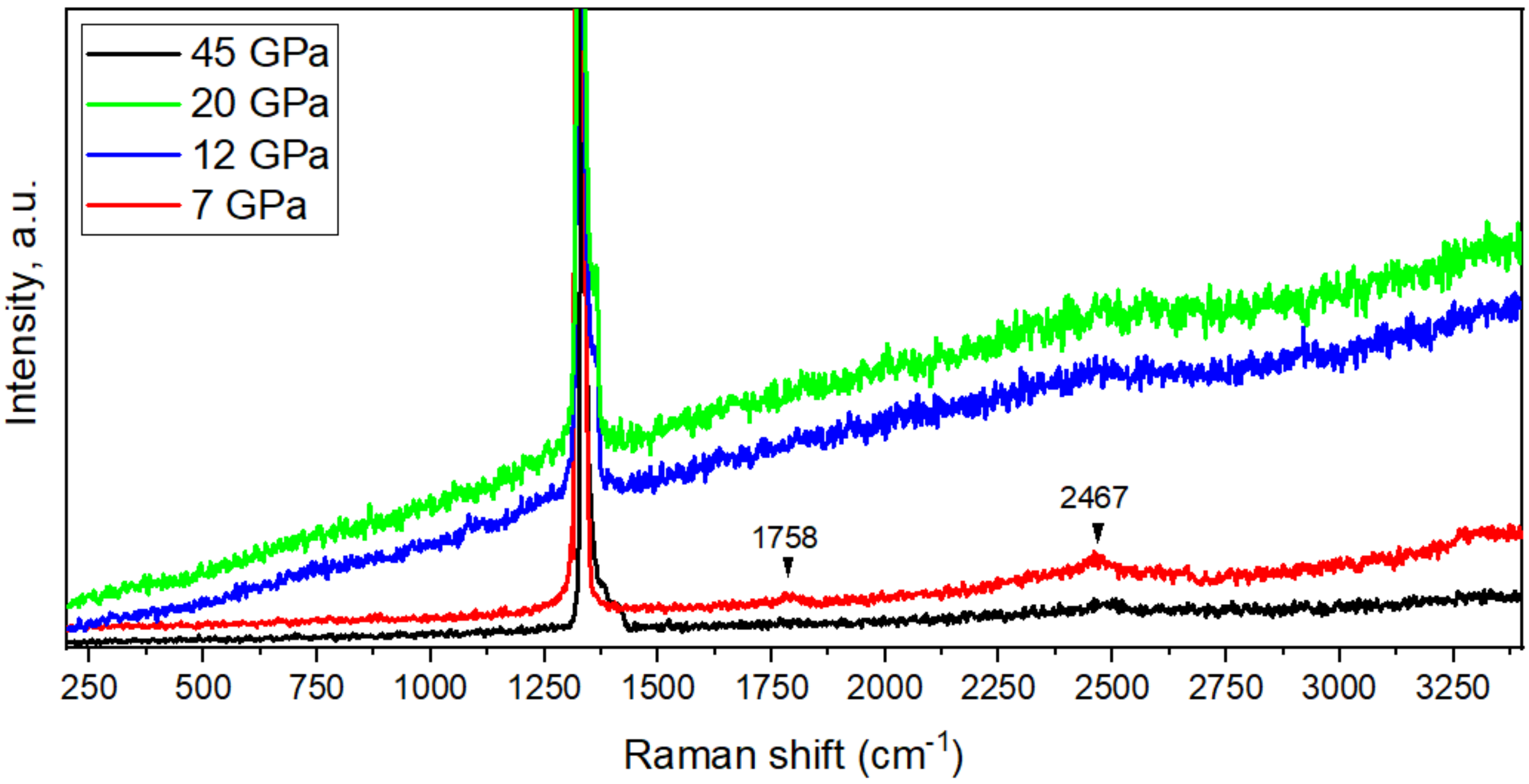

**Figure S11.** Raman spectra of the $TcH_x$ samples taken at different pressures in the range of 250–3250 $cm^{-1}$. There are weak signals (1758, 2467 $cm^{-1}$) from products of decomposition of ammonia borane (AB) and from AB itself.

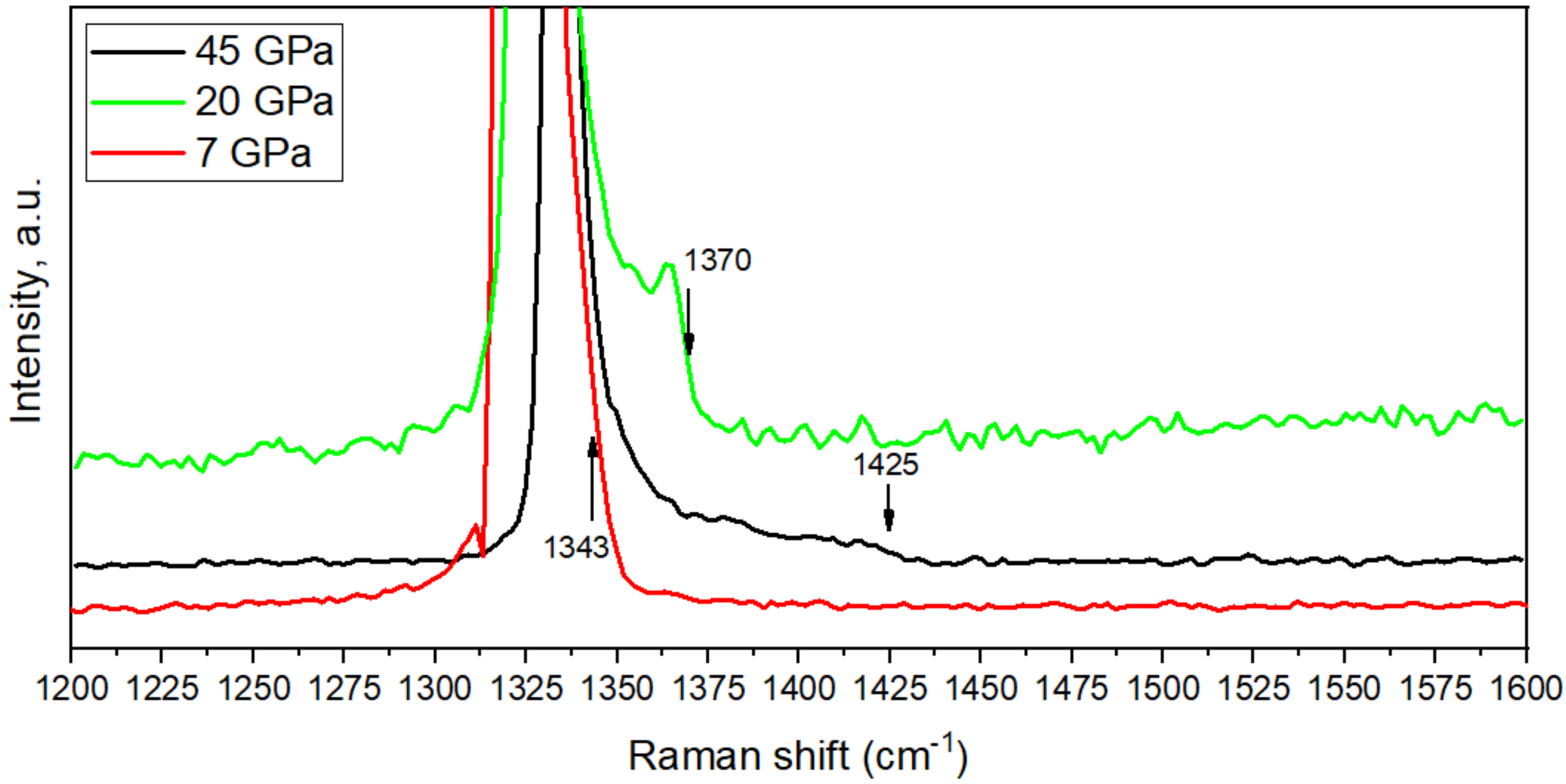

**Figure S12.** Raman spectra of the $TcH_x$ samples taken at different pressures in the range of 1200–1600 $cm^{-1}$ (pressure determination).

## Reflection spectroscopy

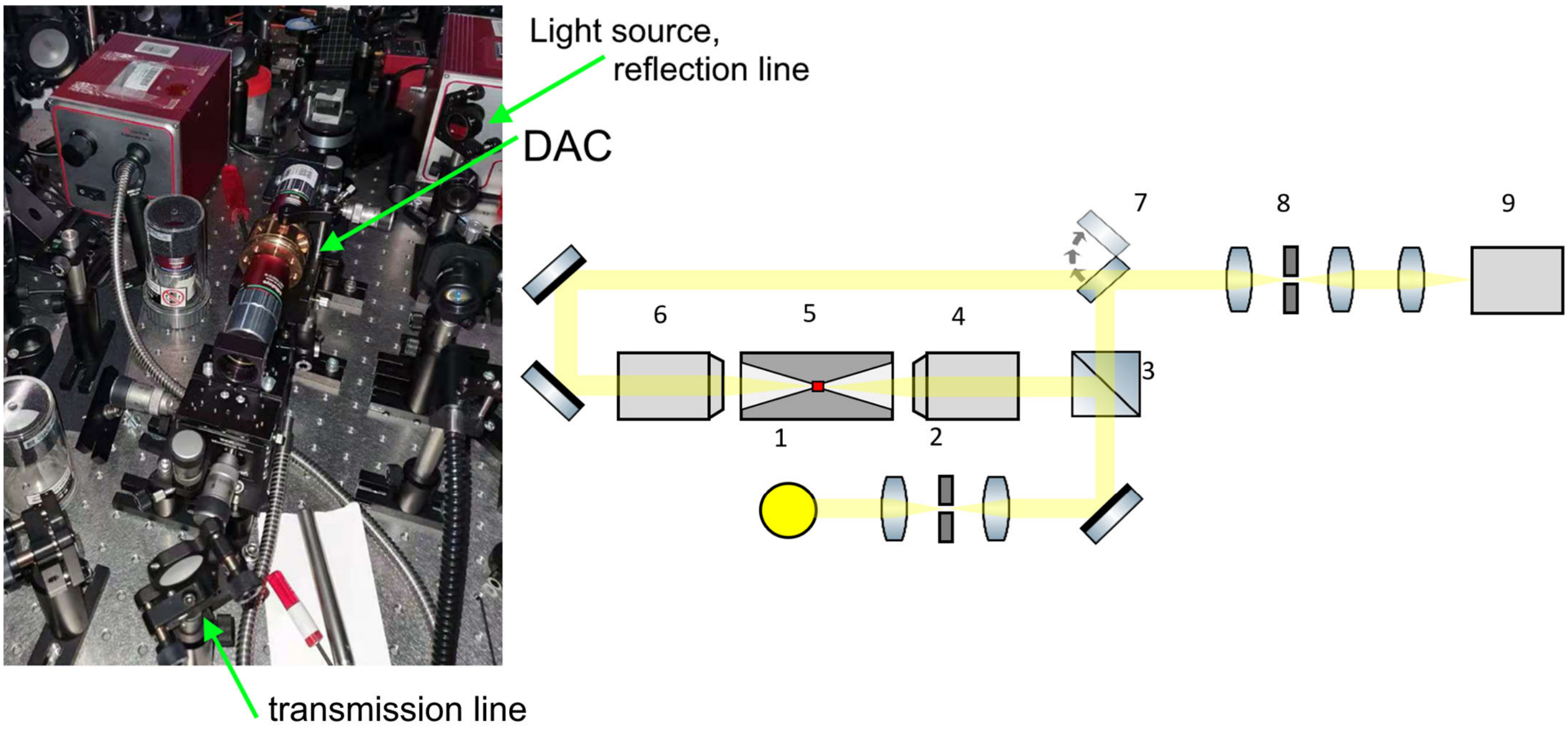

**Figure S13.** Optical scheme for transmission and reflection studying of samples in diamond anvil cells. 1 - Thorlabs OSL2B/OSL2BIR light source, 2 – a pinhole with diameter 50um, 3 – beam splitter BS013, 4 and 6 - Mitutoyo plan apo NIR 20x infinity corrected objectives, 5 – the diamond anvil sample, 7 – flip mirror, 8 – a pinhole-type spatial filter and 9 – spectrometer with CCD camera: Action SpectraPro SP-2750 and ProEM HS.

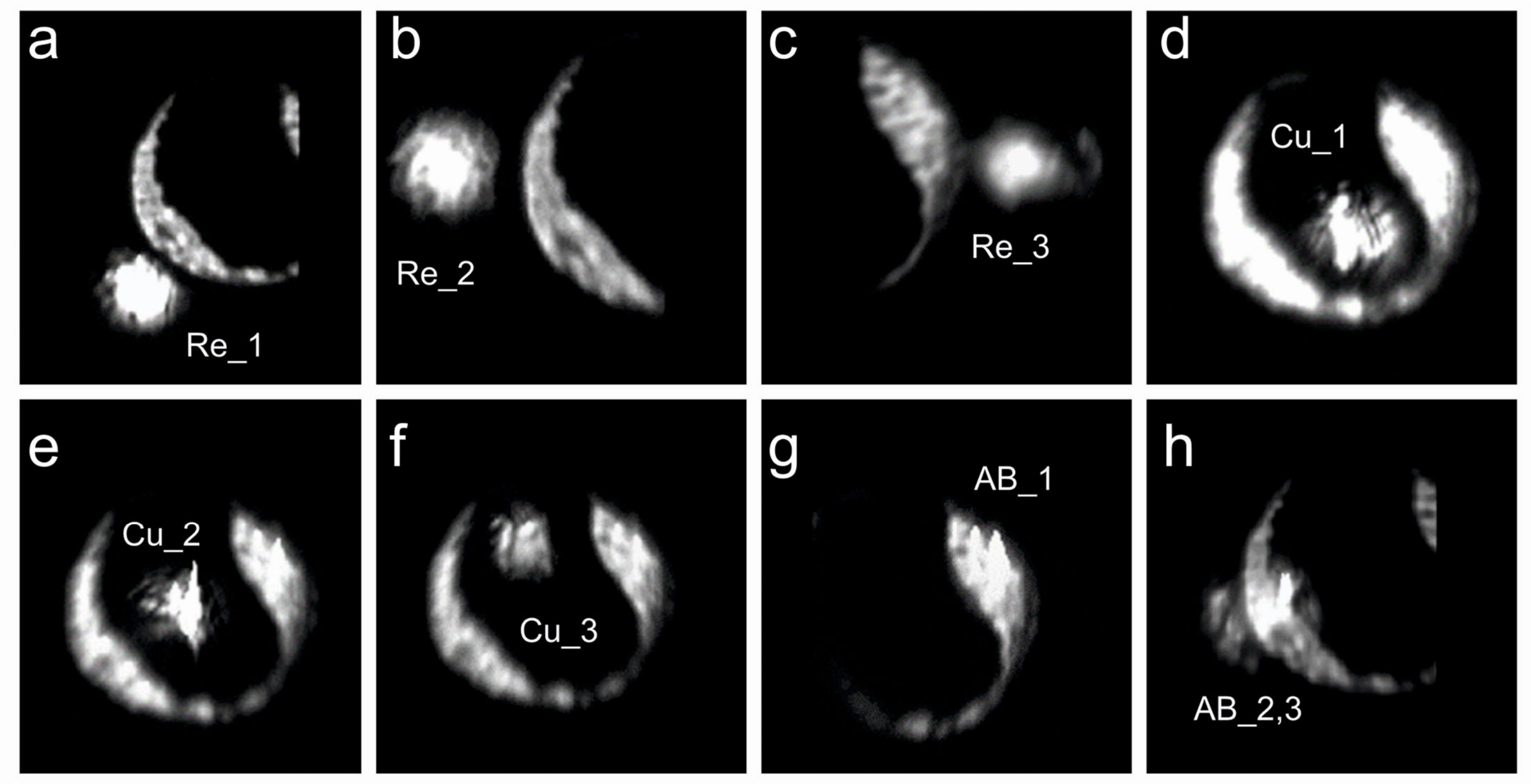

**Figure S14.** Optical photographs of the copper (Cu, d-f) sample placed in a high-pressure diamond anvil cell in a medium of ammonia borane (AB, g-h) equipped with a rhenium (Re, a-c) gasket. The bright spot of light (~15-20 μm) corresponds to the area where the measurements were made.

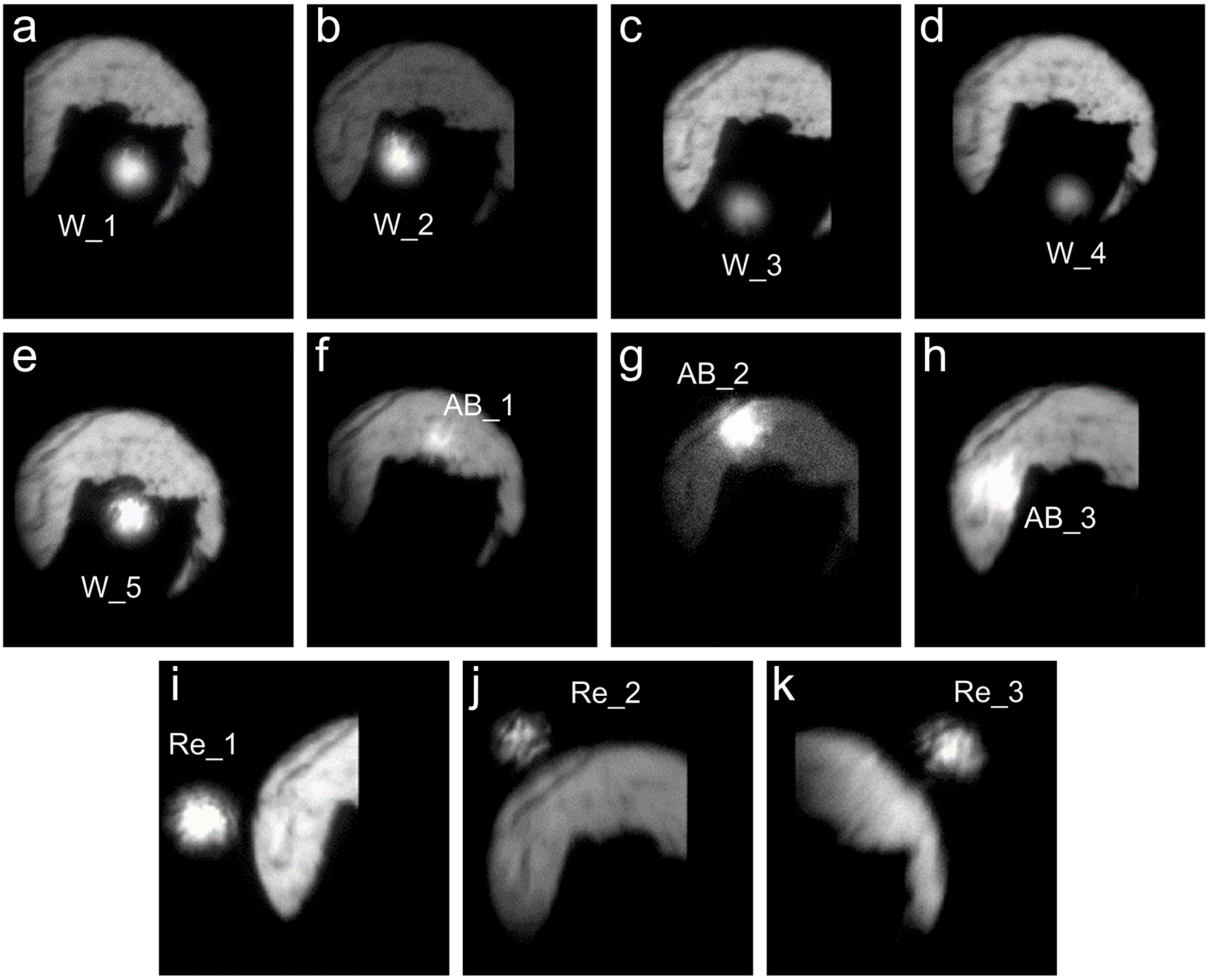

**Figure S15.** Optical photographs of a tungsten (W, a-e) sample placed in a high-pressure diamond anvil cell in a medium of ammonia borane (AB, f-h) equipped with a rhenium (Re, i-k) gasket. The bright spot of light (~15-20 μm) corresponds to the area where the measurements were made.

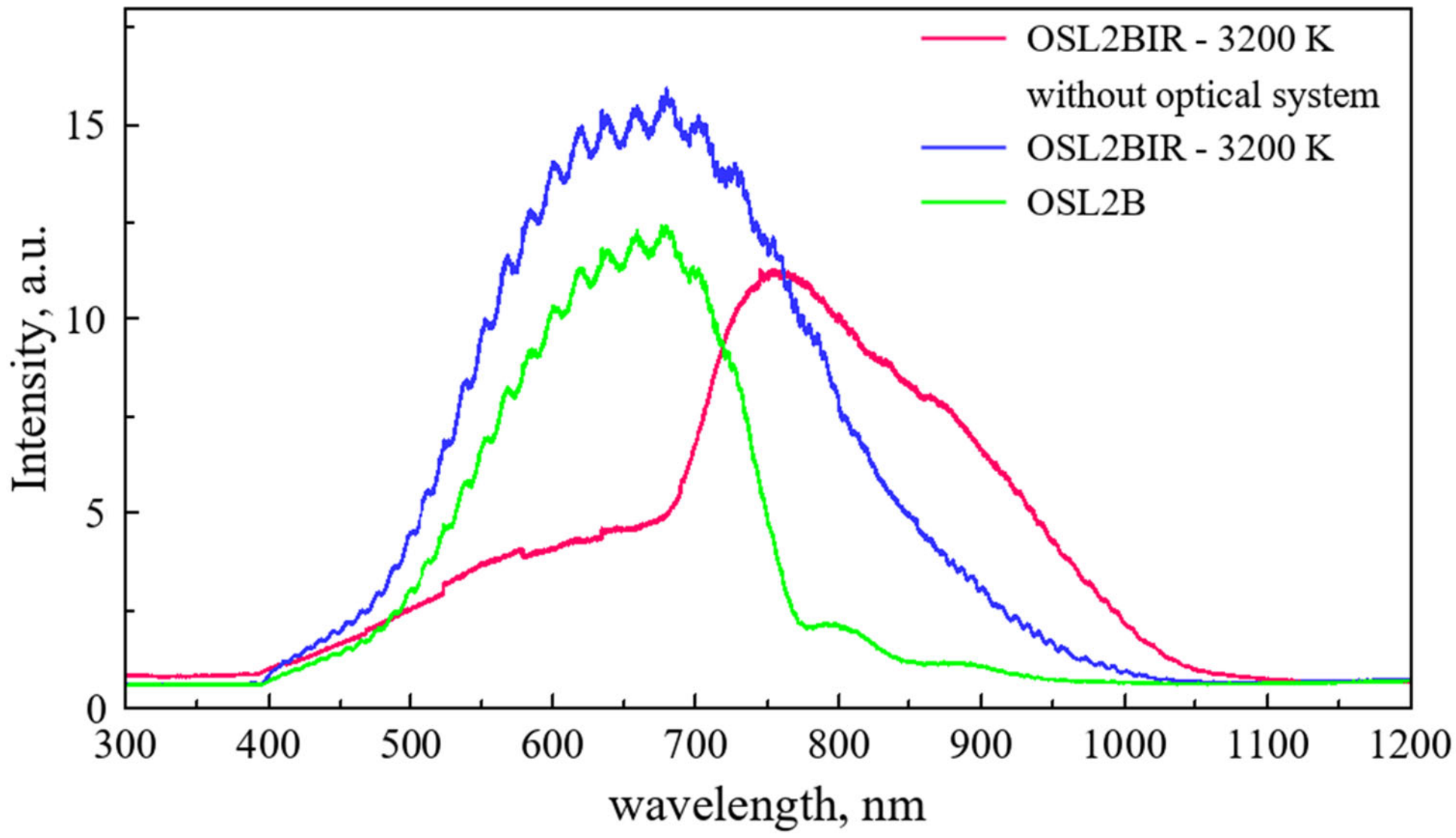

**Figure S16.** Spectral characteristics of two lamps, Thorlabs OSL2B and OSL2BIR,[29] used in the OSL2IR fiber light source with the optical scheme (interference noise is seen in the green and blue curves) and without the optical scheme (red curve, no interference noise). Oscillations on the spectra are residual interference inside optical schema. The optical setup cuts off the infrared part of the light source spectrum. The results of the reflectance measurements are irrelevant below 450 nm and  above 950 nm.

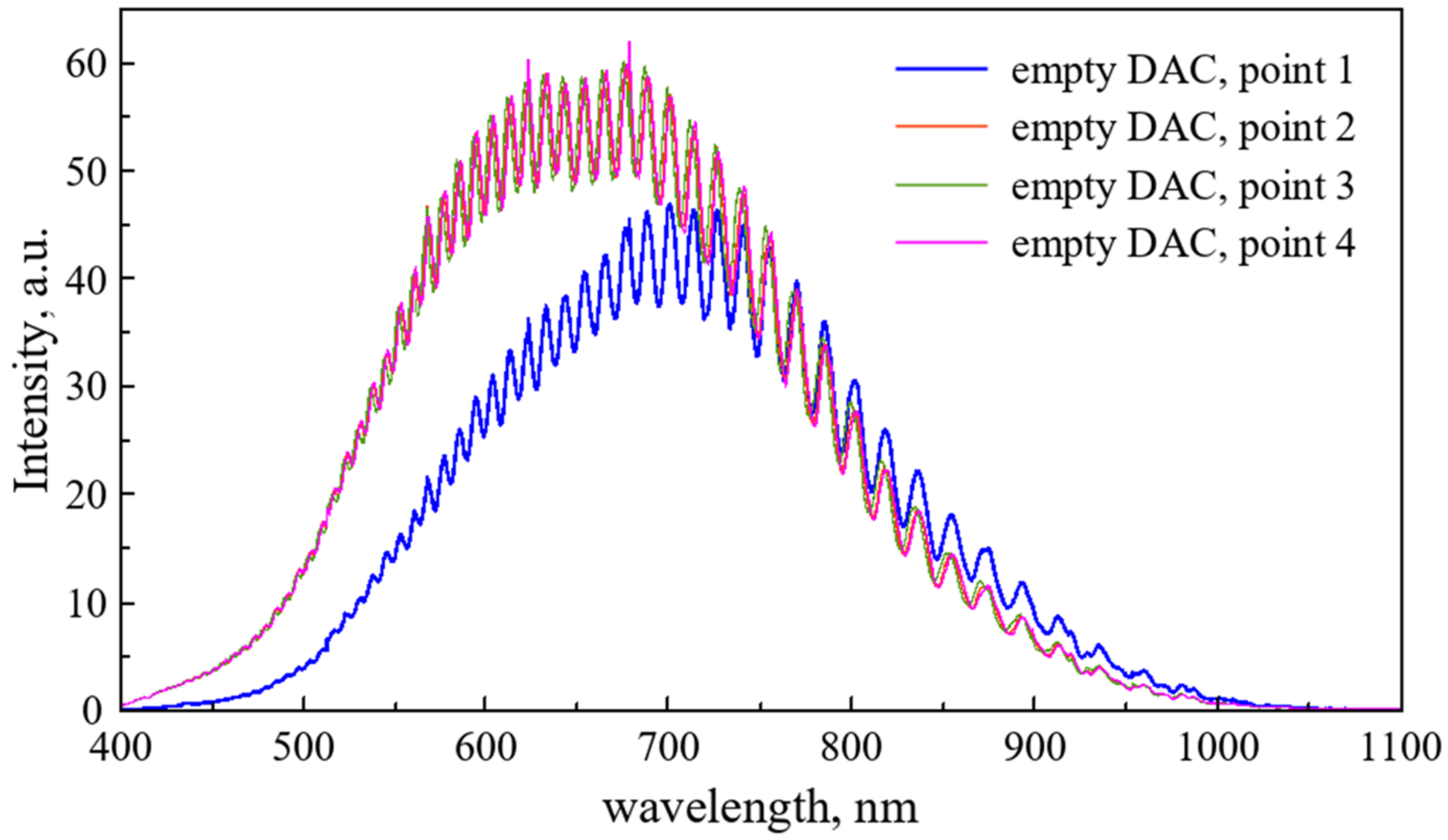

**Figure S17.** Reflection spectra of an empty diamond anvil cell measured at different points (1–4) of the diamond anvil culet. A strong interference component is present in the spectra.

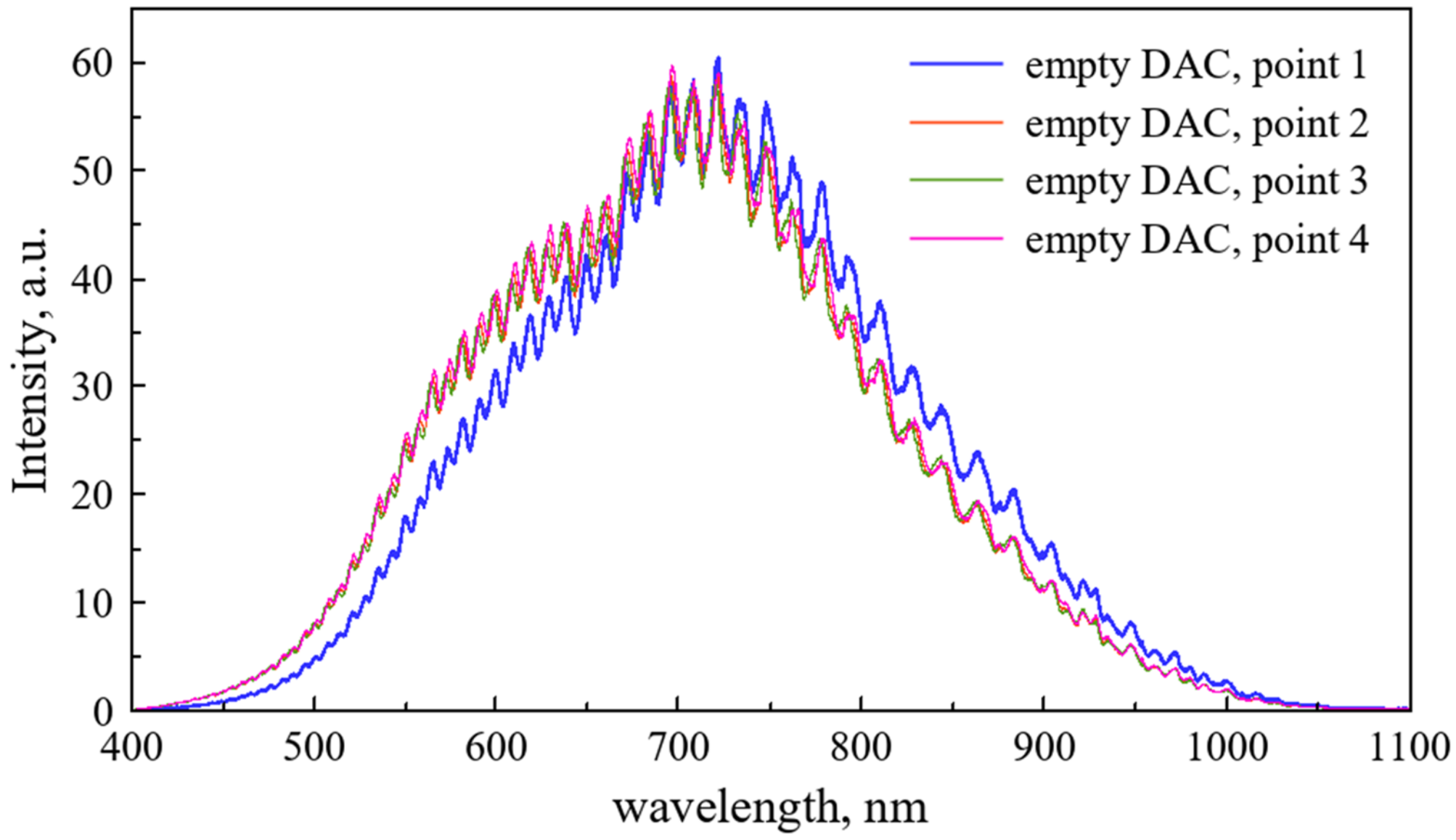

**Figure S18.** Transmission spectra of an empty diamond anvil cell measured at different points (1–4) of the diamond anvil culet. A strong interference component is present in the spectra.

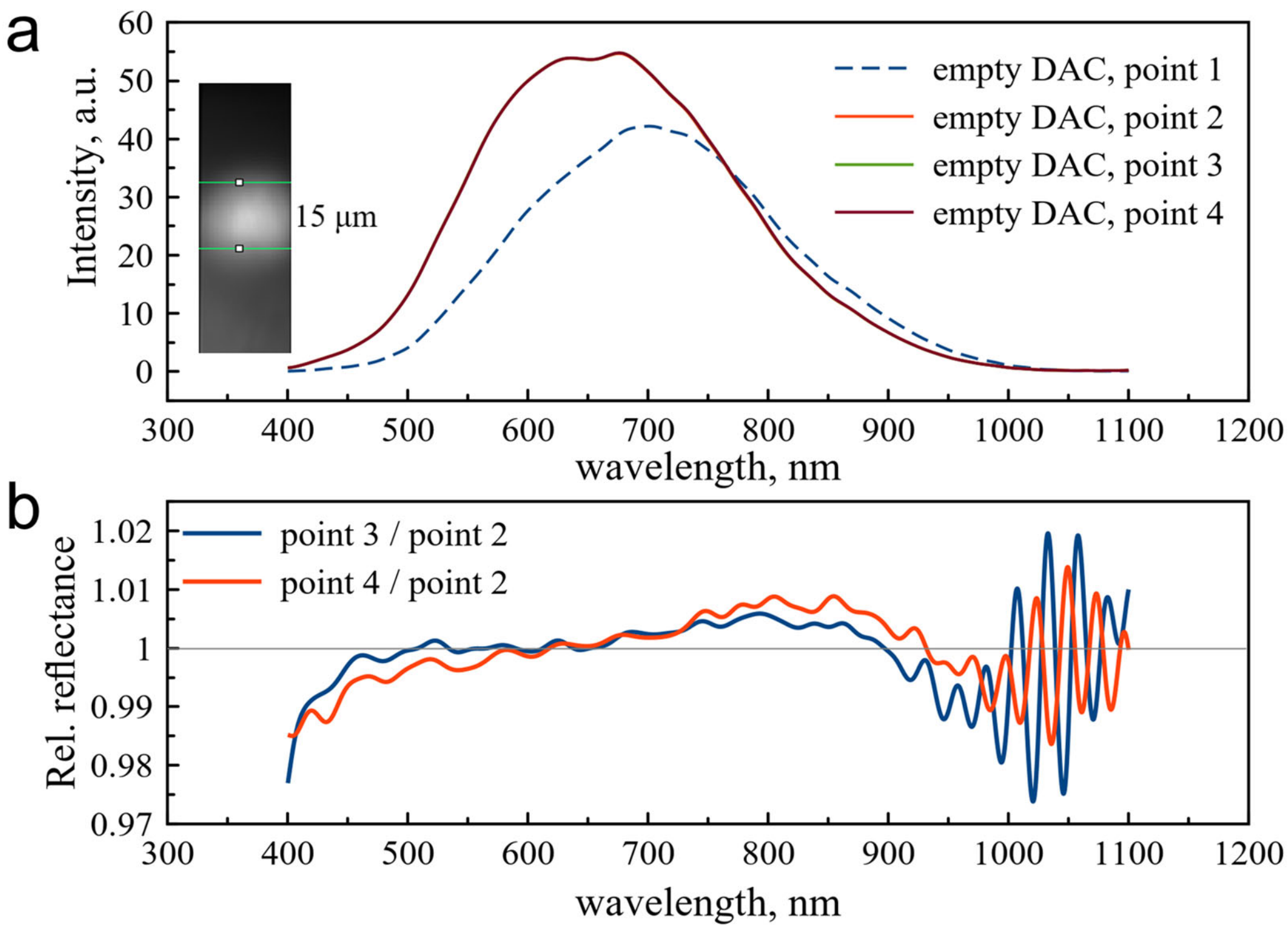

**Figure S19.** (a) Fourier-filtered reflectance spectra of the empty diamond anvil cell measured at different points (1–4, Figure S16) of the diamond anvil culet. All points except point 1 give almost identical reflectance spectra. This calibration experiment shows the importance of accumulating sufficient statistics on the reflectance spectra to draw certain conclusions. (b) Relative reflectance ($I_{sample}/I_{ref}$) of the DAC in points 3,4 with respect to point 2.

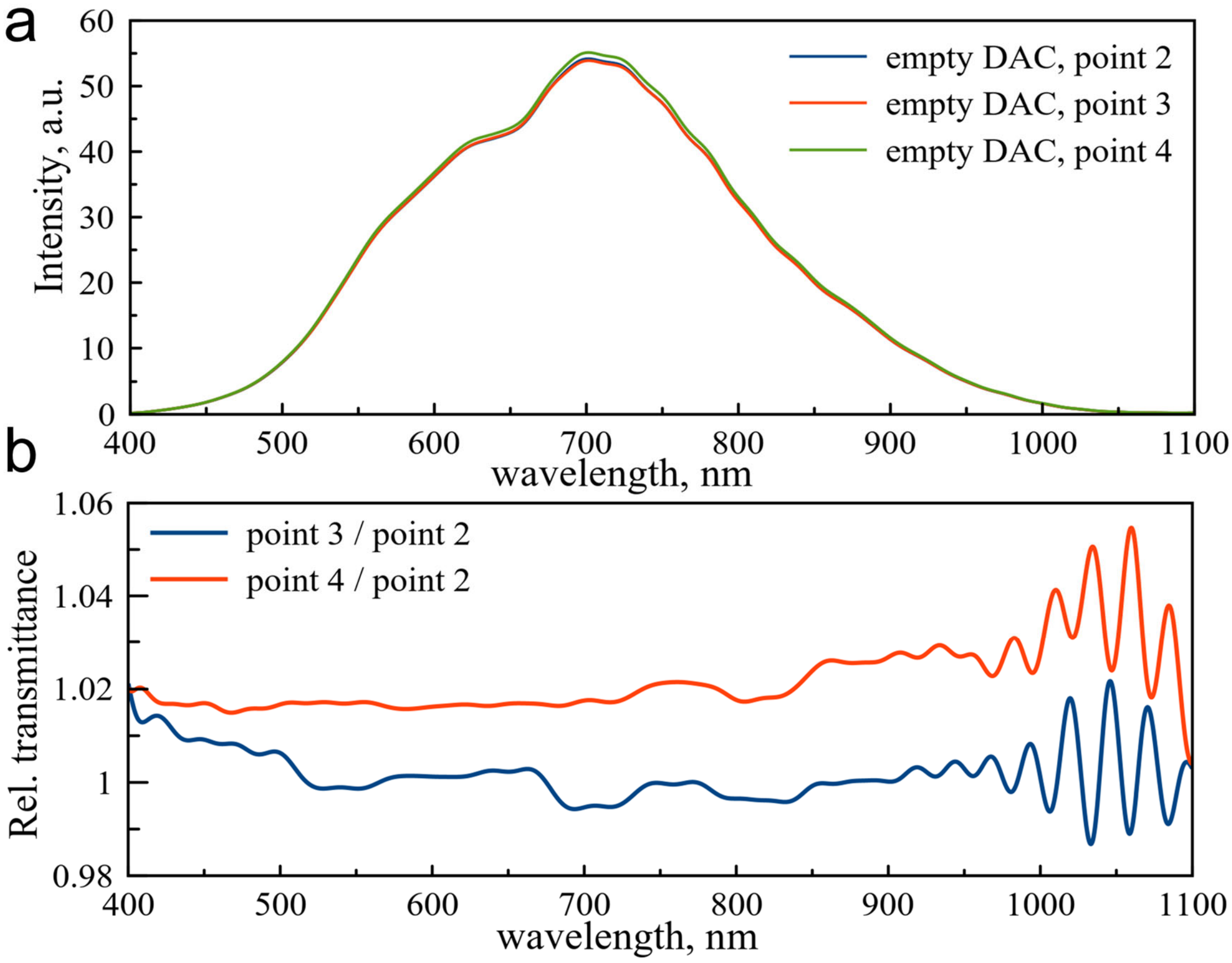

**Figure S20.** (a) Fourier-filtered transmittance spectra of the empty diamond anvil cell measured at different points (1–4, Figure S17) of the diamond anvil culet. All points except point 1 give almost identical transmittance spectra. This calibration experiment shows the importance of accumulating sufficient statistics on the transmittance spectra to draw certain conclusions. (b) Relative transmittance ($I_{sample}/I_{ref}$) of the DAC in points 3,4 with respect to point 2.

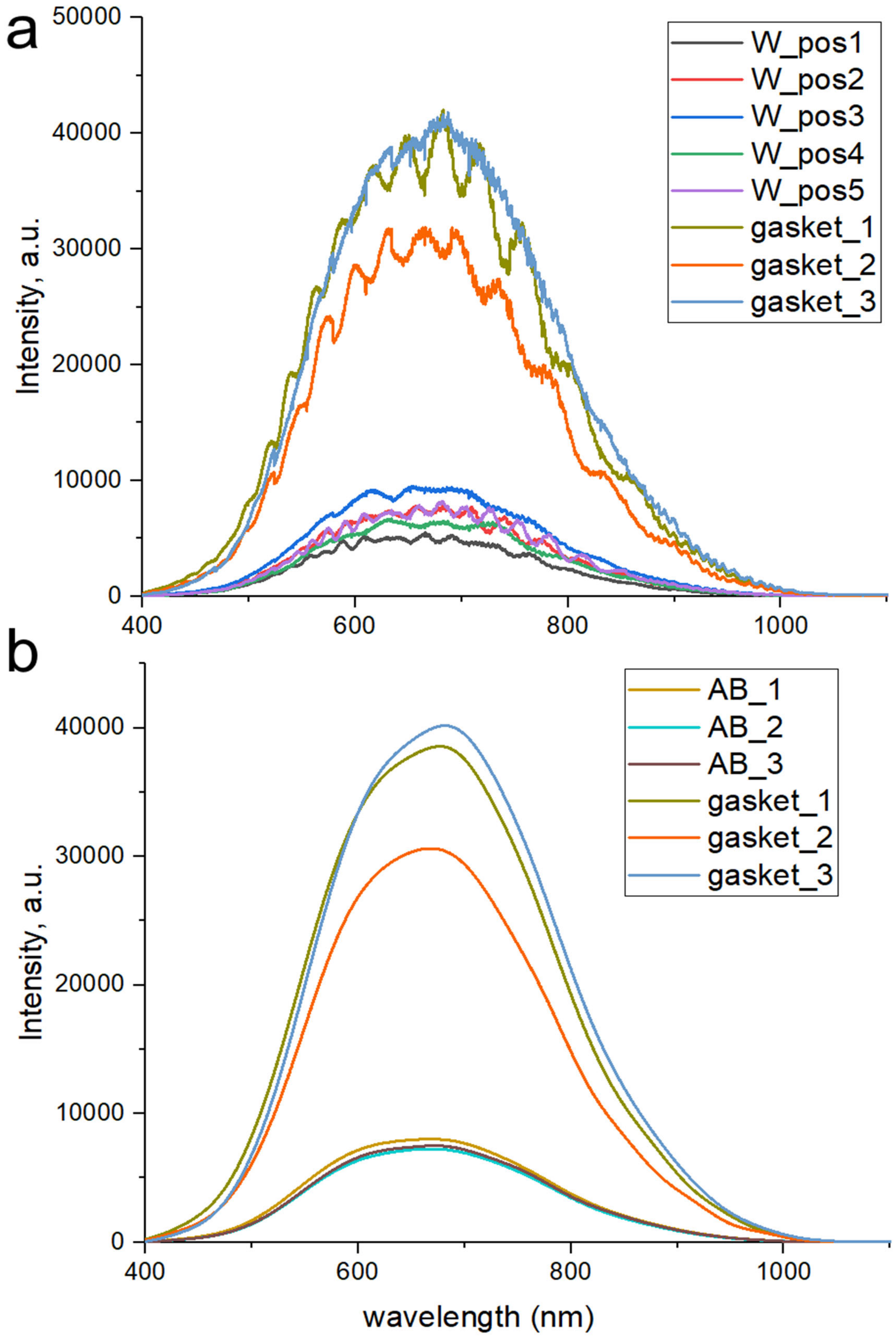

**Figure S21.** Reflectance spectra of the calibration DAC with the tungsten sample (W) in the ammonia borane medium (AB) and a rhenium gasket: (a) raw $I(\lambda)$ data, (b) Fourier-filtered spectra.

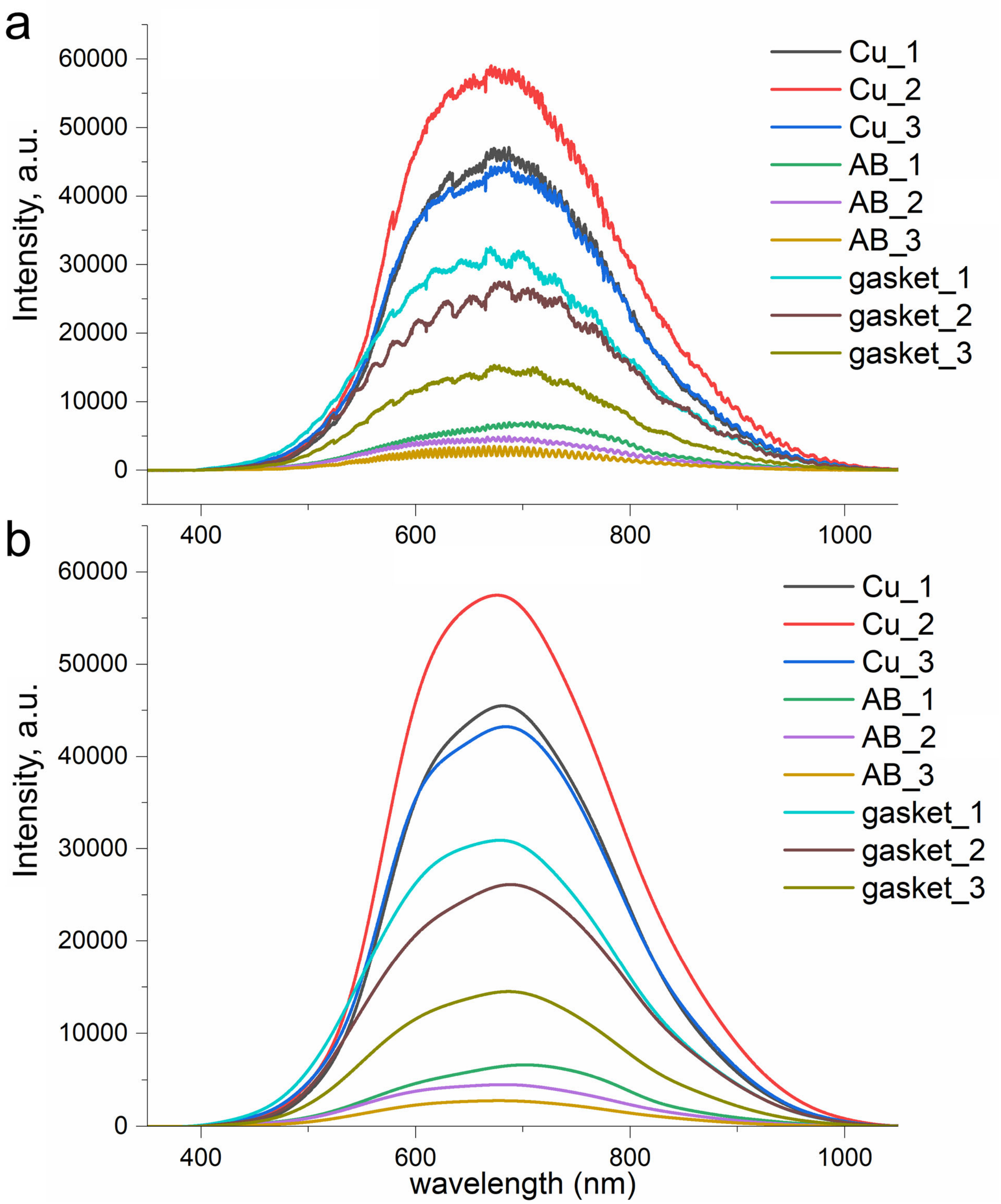

**Figure S22.** Reflectance spectra of the calibration DAC with the copper sample (Cu) in the ammonia borane medium (AB) and a rhenium gasket: (a) raw $I(\lambda)$ data, (b) Fourier-filtered spectra.

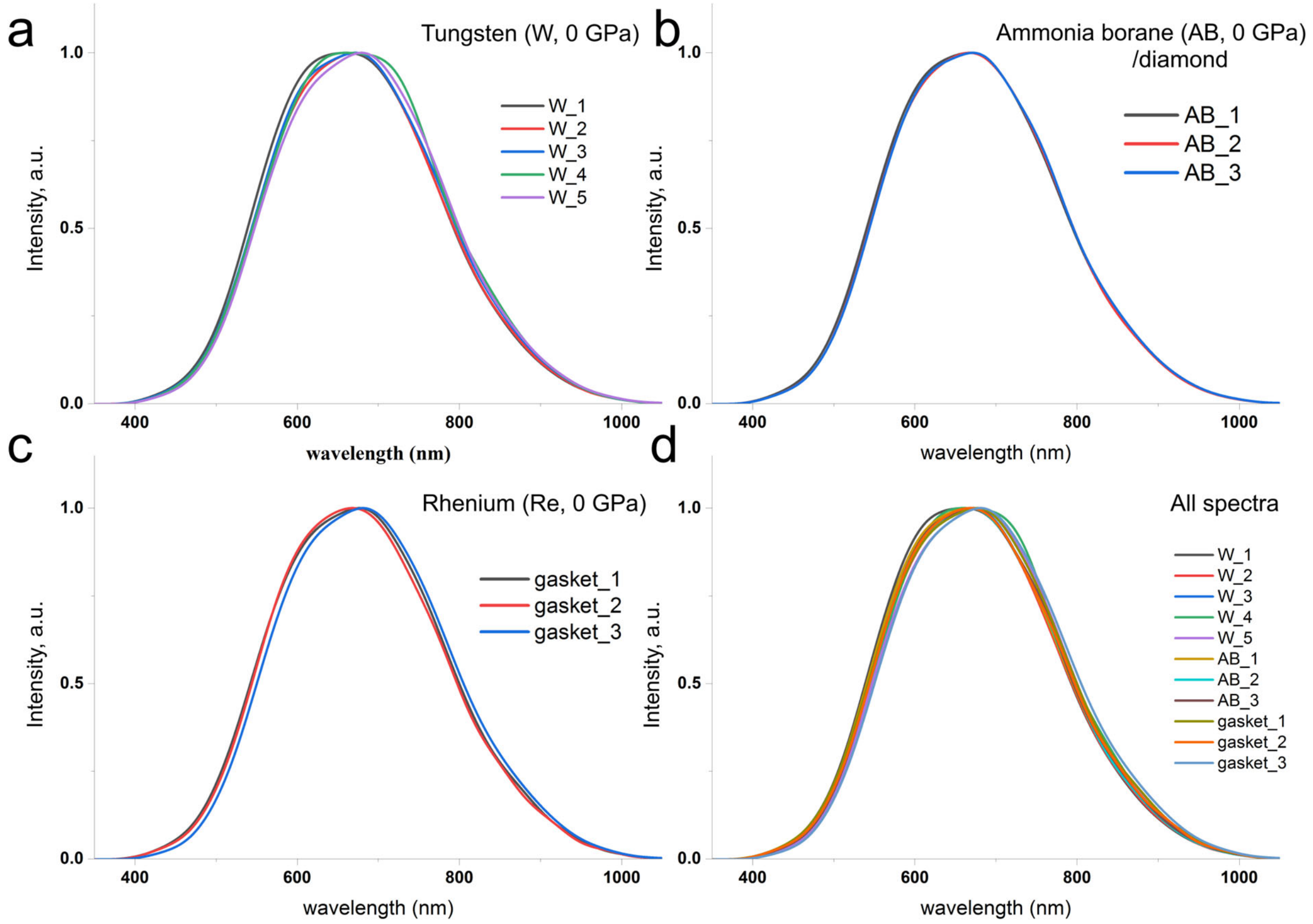

**Figure S23.** Fourier-filtered reflectance spectra, normalized to a unity, of (a) a tungsten particle loaded together with (b) ammonia borane (AB), and (c) rhenium gasket; (d) all spectra shown together.

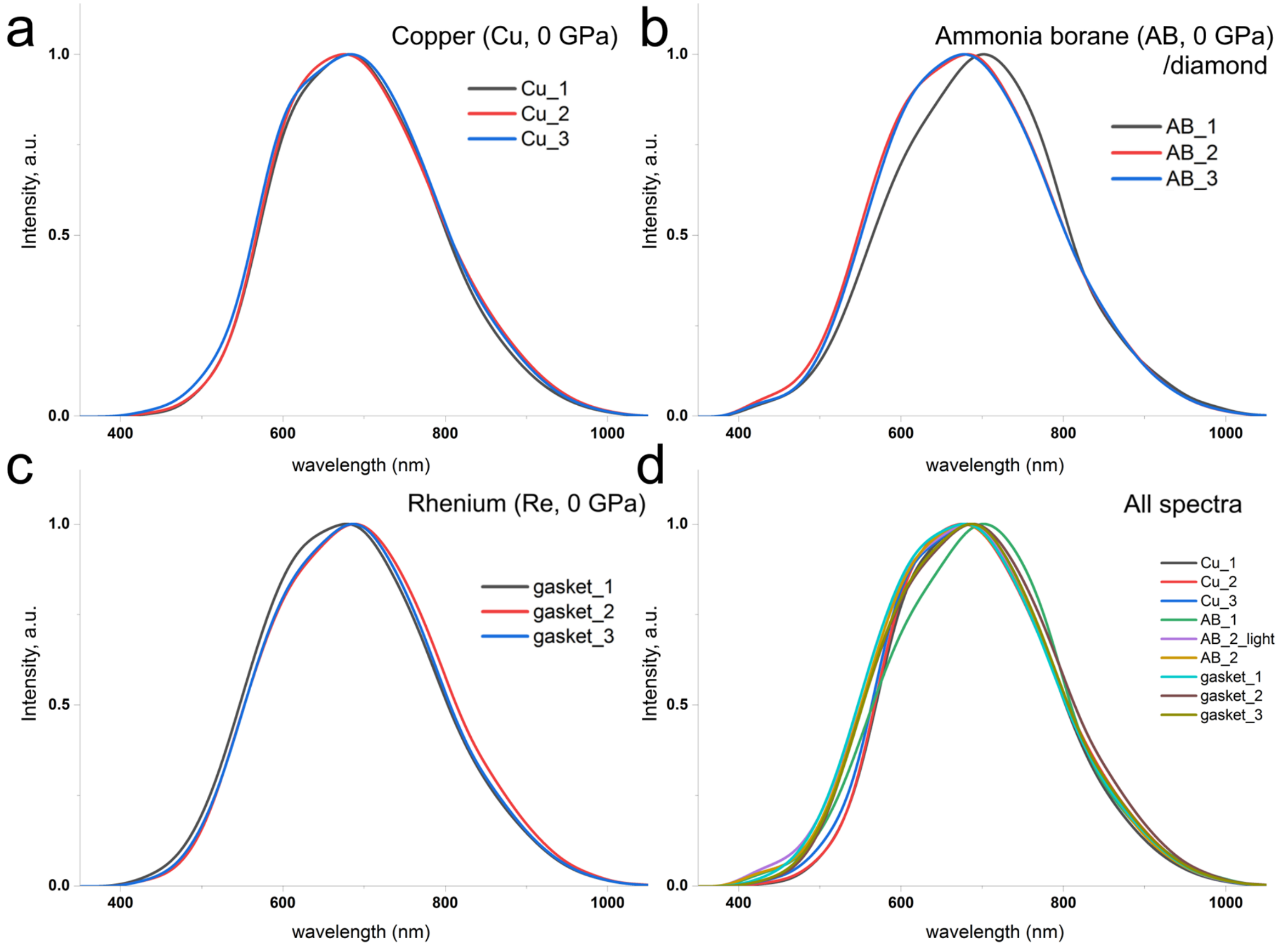

**Figure S24.** Fourier-filtered reflectance spectra, normalized to a unity, of (a) a copper particle loaded together with (b) ammonia borane (AB), and (c) rhenium gasket; (d) all spectra shown together. In panel (b), $R(\lambda)$ at point 1 obviously deviates from the behavior of the material at other points, therefore point 1 was excluded from further analysis.

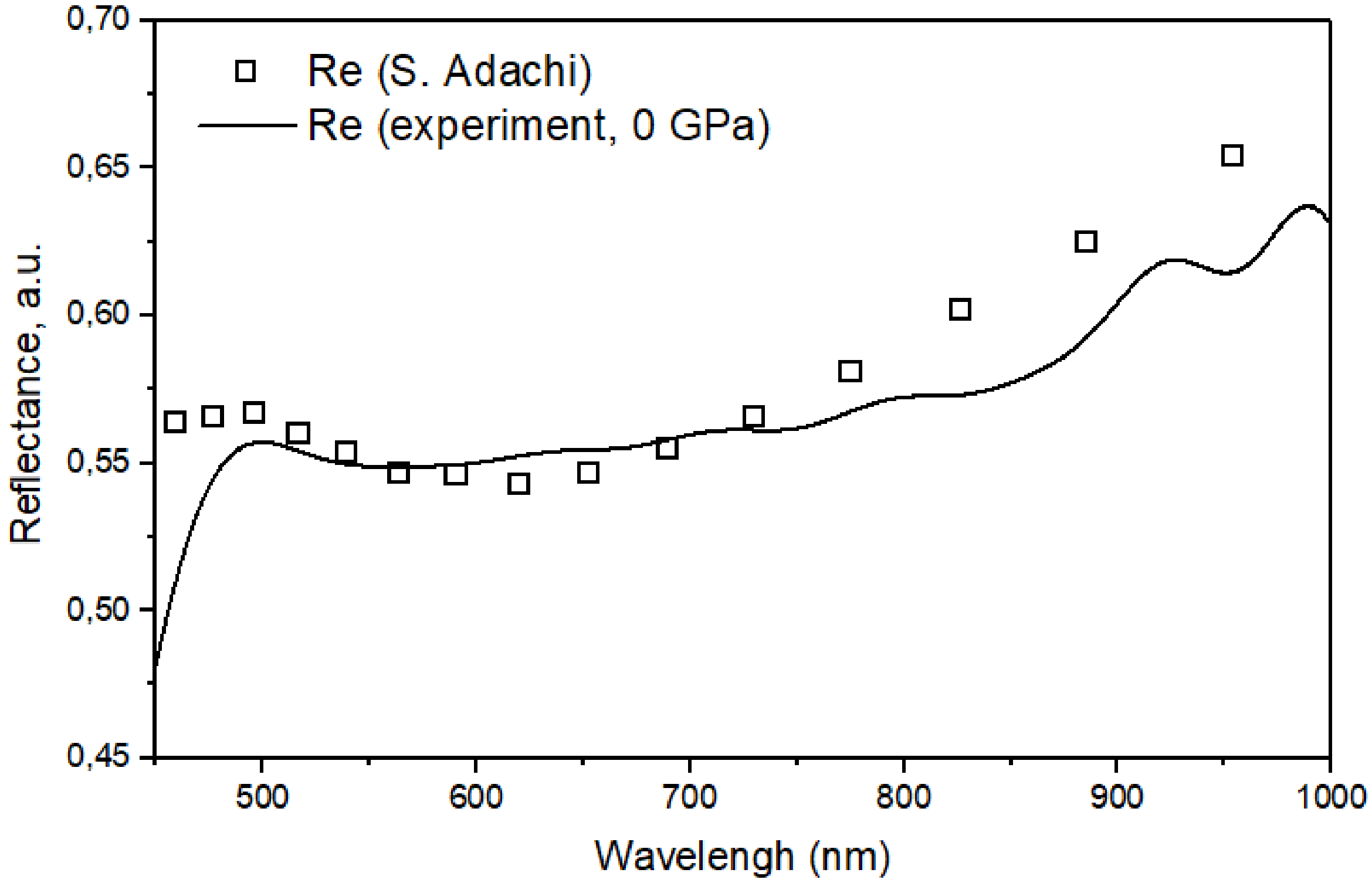

**Figure S25.** Reflectance of rhenium (Re) obtained by averaging the experiments on reflection from the gasket at 0 GPa, compared with the literature data.[30] The reflection from the diamond/ammonia borane (AB) boundary was used as the reference point. A constant of 1/1.8 was used to correct the obtained value of $I_{Re}(\lambda)/I_{AB}(\lambda)$. Upward trends at 500–600 nm and 700–1000 nm are qualitatively reproducible in the experiment.

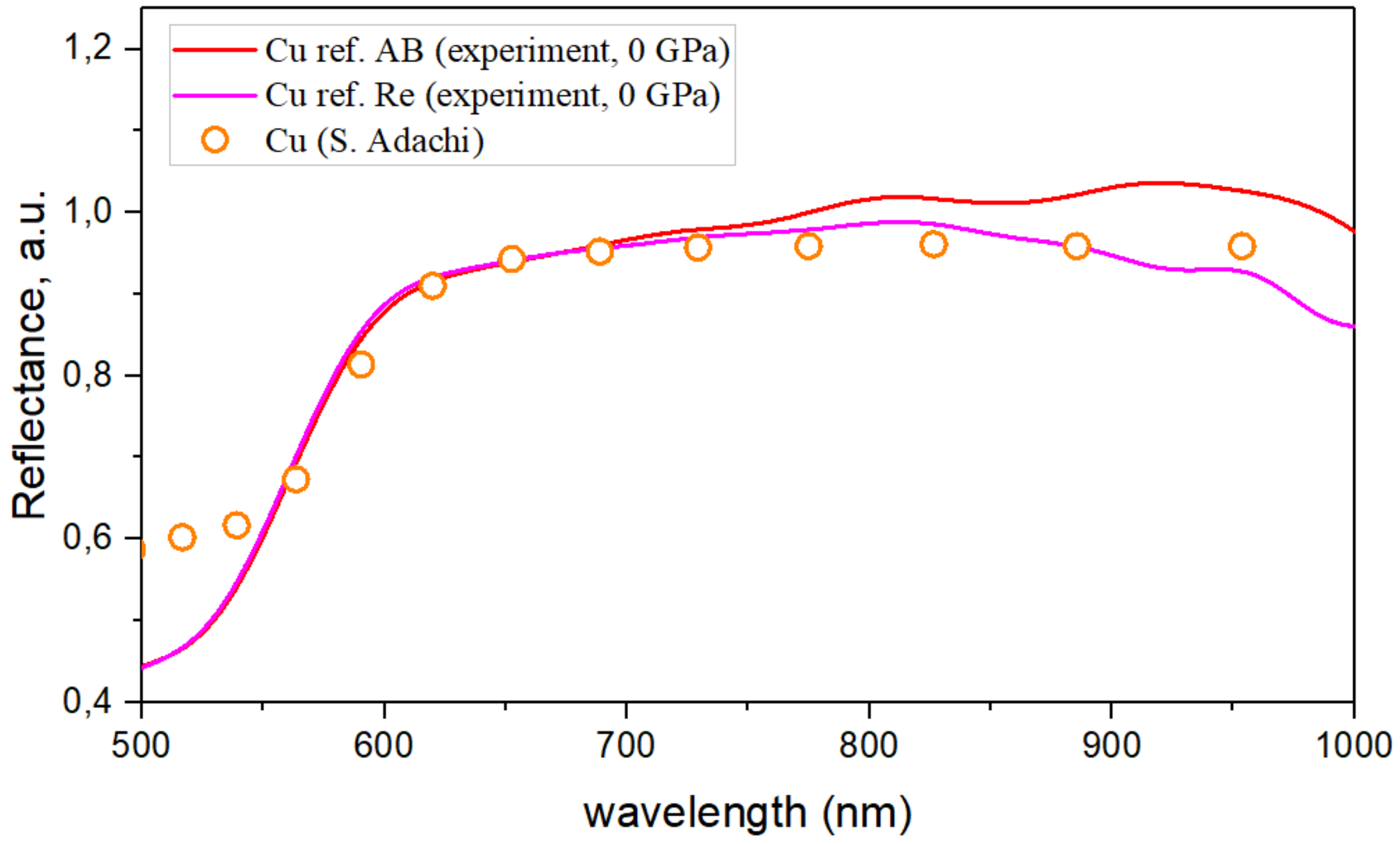

**Figure S26.** Reflectance of copper (Cu) obtained by dividing $I_{Cu}(\lambda)$ by the reflection from the AB/diamond boundary (red) and from the rhenium gasket (Re, violet), which were used as reference points. The literature data for Cu [30] is shown by orange circles. A constant of 1/1.05 was used to correct the obtained value of $I_{Cu}(\lambda)/I_{ref}(\lambda)$. A sharp decrease in the Cu reflectivity below 600 nm is well reproduced in the experiment.

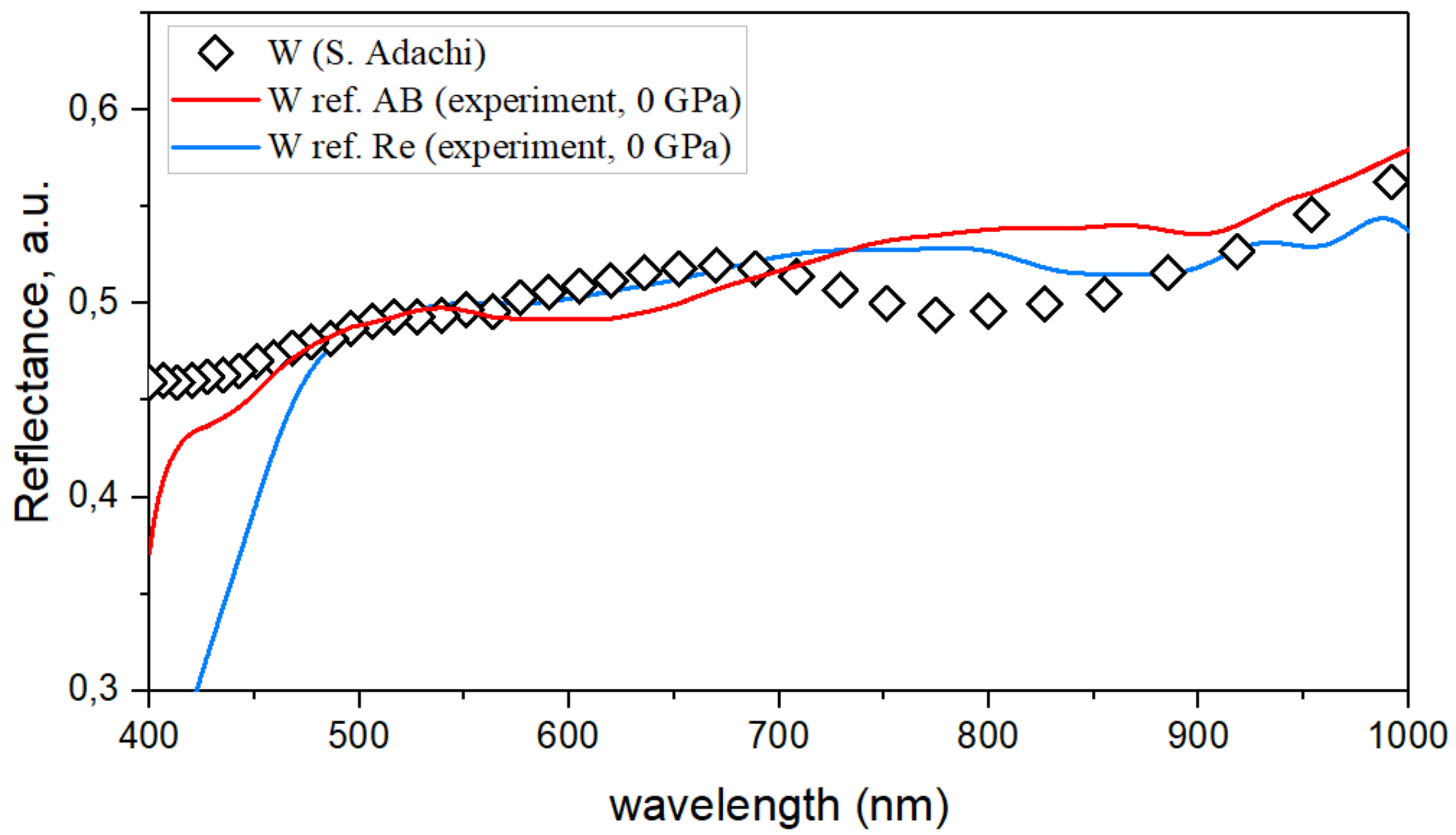

**Figure S27.** Reflectance of tungsten (W) obtained by dividing $I_W(\lambda)$ by the reflection from the AB/diamond boundary (red) and from the rhenium gasket (Re, blue), which were used as reference points. The literature data for W [30] is shown by hollow diamonds. A general trend of decreasing reflectivity of W below 0.5 when moving from 1000 nm to 400 nm is reproduced in our experiment, but we could not detect a small local decrease around 750–800 nm.